\documentclass[acmsmall]{acmart}
\AtBeginDocument{%
  \providecommand\BibTeX{{%
    \normalfont B\kern-0.5em{\scshape i\kern-0.25em b}\kern-0.8em\TeX}}}

\setcopyright{acmlicensed}
\copyrightyear{2018}
\acmYear{2018}
\acmDOI{XXXXXXX.XXXXXXX}

\acmJournal{JACM}
\acmVolume{37}
\acmNumber{4}
\acmArticle{111}
\acmMonth{8}

\newcommand{\estt}{\ensuremath{S}}
\newcommand{\smbltime}{\ensuremath{T}}
\newcommand{\wave}{\ensuremath{\Gamma}}
\newcommand{\sinu}{\ensuremath{W}}
\newcommand{\screenseglen}{\ensuremath{L}}
\newcommand{\hdrlen}{\ensuremath{D}}

\newcommand\figspacesub{-0.05in}

\newcommand\resfigw{3in}
\newcommand\resfigabw{2.5in}
\AtBeginEnvironment{bmatrix}{\setlength{\arraycolsep}{3pt}}

\renewcommand\footnotetextcopyrightpermission[1]{} 
\setcopyright{none}
\begin{document}

\title{Understanding Long Range-Frequency Hopping Spread Spectrum (LR-FHSS) with Real-World Packet Traces}

\author{Jumana Bukhari and Zhenghao Zhang}
\email{{jbukhari,zzhang2}@fsu.edu}
\affiliation{%
  \institution{Florida State University}
  \city{Tallahassee}
  \state{Florida}
  \country{USA}
  \postcode{32306. This research work was supported by the US National Science Foundation under Grant 1910268.}
}

\begin{abstract}
Long Range-Frequency Hopping Spread Spectrum (LR-FHSS) is a new physical layer option that has been recently added to the LoRa family with the promise of achieving much higher network capacity than the previous versions of LoRa. In this paper, we present our evaluation of LR-FHSS based on real-world packet traces collected with an LR-FHSS device and a receiver we designed and implemented in software. We overcame challenges due to the lack of documentations of LR-FHSS and our study is the first of its kind that processes signals transmitted by an actual LR-FHSS device with practical issues such as frequency error. Our results show that LR-FHSS meets its expectations in communication range and network capacity. We also propose customized methods for LR-FHSS that improve its performance significantly, allowing our receiver to achieve higher network capacity than those reported earlier.
\end{abstract}

\ccsdesc[500]{Networks~Network experimentation}
\ccsdesc[500]{Networks~Network simulations}




\maketitle

\thispagestyle{empty}
\pagestyle{plain}

\section{Introduction}

LoRa~\cite{RefLoRaWan} in recent years has emerged as one of the strongest wireless technologies for the connections of IoT devices to gateways potentially over long distances. Recently in November 2020, Semtech, the company behind LoRa, announced a new Physical layer (PHY) option in the LoRa family, namely, Long Range-Frequency Hopping Spread Spectrum (LR-FHSS)~\cite{LRFHSSnews}. LR-FHSS is completely different from the traditional Chirp Spread Spectrum (CSS) modulation in LoRa and is expected to achieve higher network capacity and support satellite communications in addition to terrestrial communications. Since the announcement, LR-FHSS has gathered growing interest~\cite{9422331,10004576,9653679,9825690,9799790,9691385,9842830}. 
However, as LR-FHSS was introduced only recently, there is a lack of technical documentations and open-source resources. Most of the existing work thus rely on mathematical analysis or simulations with certain simplifying assumptions~\cite{9422331,10004576,9653679,9825690,9799790}; others list it as future work~\cite{9691385,9842830}. Therefore, it would be beneficial to understand the complete details of LR-FHSS and conduct tests with signals from actual LR-FHSS devices, so that the network performance can evaluated when real-world issues, such as timing error, frequency estimation error, limits in error correction, and effectiveness of Successive Interference Cancellation (SIC), are taken into account. 

In this paper, we present our study of LR-FHSS based on real-world packet traces. To the best of our knowledge, our study is the first of its kind that decodes actual signals transmitted by an LR-FHSS device. We used {\em SX1261}~\cite{LRFHSSdevice} as the transmitter, which is a commodity LR-FHSS development kit made available recently, and a USRP B210~\cite{USRPB210} as the receiver. We collected traces of 1000 randomly generated packets, where each trace consists of baseband samples taken by the USRP during the transmission of the packet. We also wrote software that can decode the trace correctly, i.e., convert the received baseband waveform into bits and pass the Cyclic Redundancy Check (CRC). The design and implementation of our receiver have been demonstrated with real-world experiments in the POWDER wireless platform~\cite{RefPowder}, which is an open platform with radios that can be controlled remotely. We used 4 radios as LR-FHSS nodes and one radio as the gateway and decoded almost all packets from the nodes when all nodes transmitted packets simultaneously almost non-stop. For a quantitative evaluation in networks of larger sizes, we relied on trace-driven simulations. For example, we added noise to the signal to find the Packet Receiving Ratio (PRR) as a function of the Signal to Noise Ratio (SNR), from which the expected communication distance could be revealed. We also mixed the signals of a large number of packets into a synthesized trace, which emulates the actual signal received by the gateway when nodes in the network transmit packets at random times. From the number of packets that were decoded correctly by our receiver, the network capacity could be revealed. We have made our code and data set publicly available at~\cite{LR-FHSS-receiver}.

Our finding is, in short, that LR-FHSS mostly lives up to its expectations. That is, LR-FHSS signals can be decoded at low SNRs, such as -20 dB, which is comparable to the traditional CSS modulation~\cite{semtechperformance} and therefore should achieve a similar communication distance. Also, as suggested earlier in~\cite{9422331,9653679,semtechperformance}, the network capacity with LR-FHSS is indeed much higher than that with CSS. It should be noted that the capacity with our receiver is actually higher than those reported in~\cite{9422331,9653679,semtechperformance} mostly due to a few customized methods we designed for LR-FHSS. We also confirm that LR-FHSS achieves similar network capacities in terrestrial and satellite channels.  

We overcame a number of challenges in this study, which were mostly due to the lack of documentation of LR-FHSS. To decode the baseband waveform, complete information about signal modulation and packet format is needed. When we started this work, while various aspects of LR-FHSS have been discussed in sources such as~\cite{LoRaPara,9422331,sx126xdriver}, the information was not complete. It took us a bit of reverse engineering and some luck to be able to understand all details of LR-FHSS. The receiver design was also a challenge, because at the time, there were no available information on key issues such as how to detect a packet, perform symbol-level synchronization, and estimate the signal frequency. On the other hand, a poorly designed receiver results in poor performance and does not do justice for LR-FHSS. We were able to design every component of the receiver from scratch, as well as proposing new methods, namely, customized erasure decoding and SIC, which were not mentioned in earlier studies but significantly improve the performance of LR-FHSS.

We are ready to admit that, to evaluate LR-FHSS, the ideal situation would be to deploy LR-FHSS nodes in a large area and record the signals from the nodes. While we have demonstrated our receiver design and implementation with a small network in the POWDER platform~\cite{RefPowder}, our main approach was the trace-driven simulation due to a few practical constraints we could not overcome. First, as LR-FHSS achieves high capacity, to probe the actual limit of LR-FHSS, the number of nodes is large, i.e., over 60, which is beyond our current capability. Second, as LR-FHSS was designed to support satellite communications, to evaluate the performance LR-FHSS, the access of at least one satellite is needed, which is even more challenging. Our trace-driven simulation does allow us to evaluate LR-FHSS when the real-world issues mentioned earlier are taken into account because the signals are from a real LR-FHSS transmitter. As our packet traces were collected under ideal conditions, i.e., with a strong line-of-sight path, the traces can be regarded as clean signals from a LR-FHSS transmitter, making them versatile for various kinds of purposes. For example, in our experiments in the POWDER platform~\cite{RefPowder}, the radios could emulate LR-FHSS nodes because they could simply play the packet traces. Also, to evaluate LR-FHSS in various channels, the packet traces can be processed by widely accepted channel models to emulate the received signal in such channels, where the models can be the ETU model~\cite{Ref3GPPTS36101,Ref3GPPTS36104} for the terrestrial channel and the NTN model~\cite{3GPPTR38811,3GPPTR38901,3GPPTR3821,ITURP68111,Refmatlabntt} for the satellite channel. As we have made our source code and data set public for the community at~\cite{LR-FHSS-receiver}, our packet traces will also facilitate future work on LR-FHSS by other researchers to suit their specific needs. It may be worth mentioning that the network capacity evaluation by Semtech itself~\cite{semtechperformance} was also based on simulations.

To summarize, our contributions in this paper include: 1) complete documentation of LR-FHSS, source code and packet traces with free access~\cite{LR-FHSS-receiver}, 2) a receiver design that achieves higher performance than those reported in earlier studies including that by Semtech itself~\cite{semtechperformance}, and 3) an in-depth study of LR-FHSS that reveals the expected performance of LR-FHSS in terrestrial and satellite networks with real-world issues considered.  
 
The rest of the paper is organized as follows. 
Section~\ref{sec:related} discusses related work.
Section~\ref{sec:background} gives a short description of LR-FHSS.
Section~\ref{sec:tracecollect} explains the trace collection.
Section~\ref{sec:receiver} explains our receiver design.
Section~\ref{sec:powderexp} experimentally demonstrates our receiver.
Section~\ref{sec:evaluation} evaluates LR-FHSS.
Section~\ref{sec:conclusion} concludes the paper.

\section{Related Work}
\label{sec:related}

Many LPWAN technologies have emerged in recent years, such as LoRa~\cite{RefLoRaWan}, Sigfox~\cite{RefSigfox}, RPMA~\cite{RefRPMA,RPMAPatent}, NB-IoT~\cite{RefNBIoT,DBLP:journals/access/ChenMHH17,DBLP:journals/cm/WangLAGSBBS17}, etc., among which LoRa has attracted significant attention and has been deployed world-wide. The traditional physical layer of LoRa is based on the Chirp Spread Spectrum (CSS), which has been extensively studied~\cite{DBLP:conf/sigcomm/EletrebyZKY17,8888038,9178997,DBLP:conf/infocom/TongXW20,DBLP:conf/mobisys/TongWL20,9259397,9488695,9651985,PCube,DBLP:conf/sigcomm/ShahidPCBK21}. LR-FHSS is completely different from CSS because LR-FHSS signals occupy a very narrow bandwidth while CSS signals are chirps that occupy the entire system bandwidth. LR-FHSS has been branded by Semtech for supporting both satellite and terrestrial communications. Our work is different from studies on satellite communications based on the traditional CSS modulation, such as~\cite{9887761,10059811}. 

There have been increasing interests in LR-FHSS. In~\cite{9422331}, the overall design of LR-FHSS was explained and the performance analyzed with simulations, where the results show that the network capacity with LR-FHSS is more than an order or magnitude higher than that with the traditional CSS modulation. In~\cite{9825690}, the uplink of LoRa to satellite was studied, including the CSS and LR-FHSS modulations, where the results were obtained with simulations. In~\cite{9653679}, LR-FHSS was studied theoretically and further verified with simulations. In~\cite{10004576}, a closed-form solution 
of the outage probability of LR-FHSS in satellite-based IoT networks was given. In~\cite{9799790}, connectivities of long range satellite links, including LR-FHSS links, were studied with simulations. Due to the lack of information on LR-FHSS, certain simplifying assumptions were adopted in existing studies, such as perfect timing and frequency estimation, perfect error correction, etc. On the other hand, an actual LR-FHSS receiver may not estimate the timing and frequency perfectly, and may not be able to decode a codeword correctly even if the number of symbols under collision is below a threshold. Further, Successive Interference Cancellation (SIC) has not been considered for LR-FHSS in earlier work, which is actually one of the main factors in determining the network capacity. The need of an experimental or trace-driven study has been pointed out in ~\cite{9653679}: ``the presented results should be re-validated and detailed by the follow-up studies, especially the experimental ones,'' which is the focus of our work.

Semtech itself has reported the performance of LR-FHSS in a recent application note~\cite{semtechperformance}, which includes link packet loss ratio obtained with experiments and network capacity obtained with simulations. Our work serves as an independent study which adds to the credibility of LR-FHSS. As will be discussed in Section~\ref{sec:evaluation}, the network capacity achieved with our receiver is higher than that reported in~\cite{semtechperformance}, confirming our contributions in the receiver design.  




\section{Background}
\label{sec:background}

LR-FHSS was designed to support {\em uplink} communications from  nodes to a gateway. The parameters of LR-FHSS depend on specific regions, such as the EU or the US; however, the difference is not fundamental. For example, the main difference between the EU and US is the amount of system bandwidth while the underlying signal modulations are the same. For simplicity, in this paper, the focus is on the EU region, where two data rates, denoted as DR8 and DR9, are supported. The Operating Channel Width (OCW), which is the total system bandwidth, is 137 kHz. 

\begin{figure} 
	\begin{center}
		\includegraphics[width=3in]{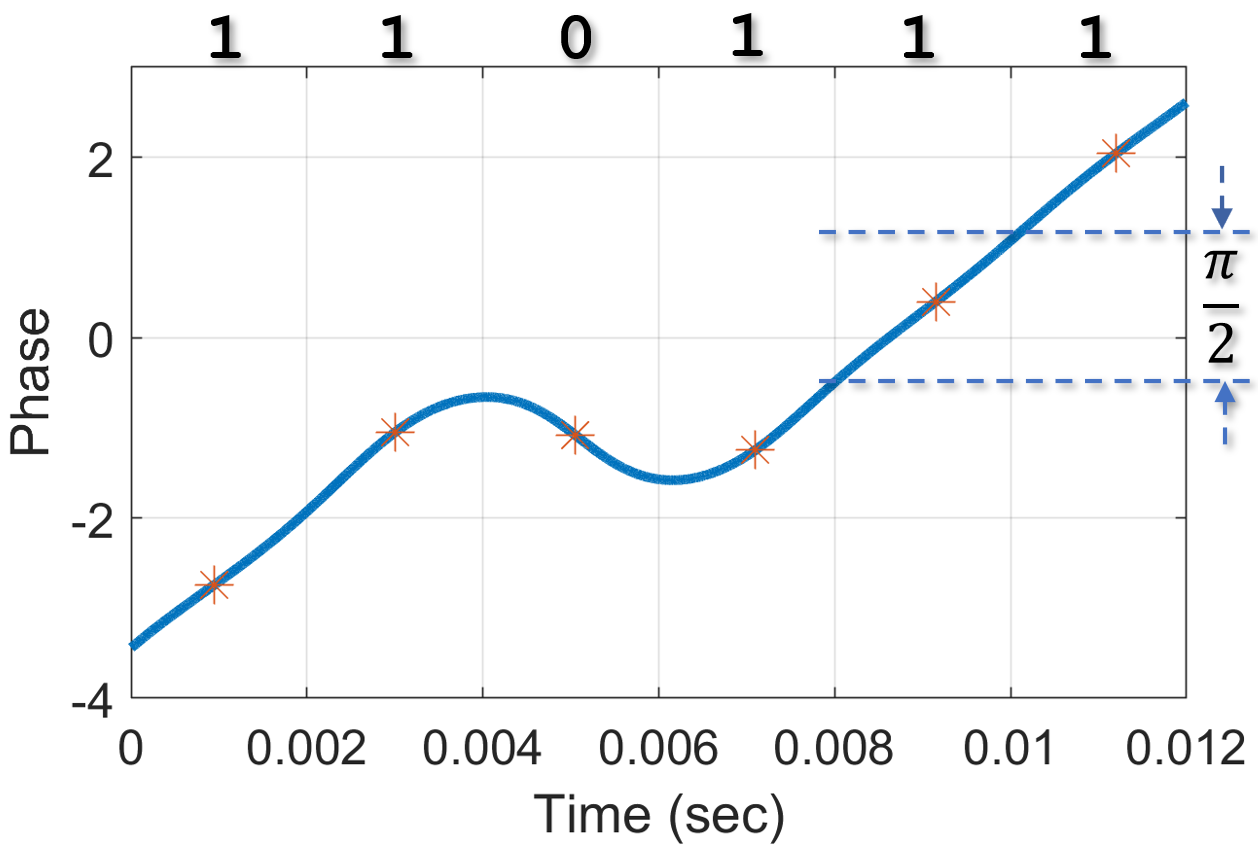} 
	\end{center}
	\vspace{\figspacesub}
	\caption{An example of GMSK modulation. }
	\label{fig:ex_GMSK}
\end{figure}

\begin{figure} 
	\begin{center}
		\includegraphics[width=4in]{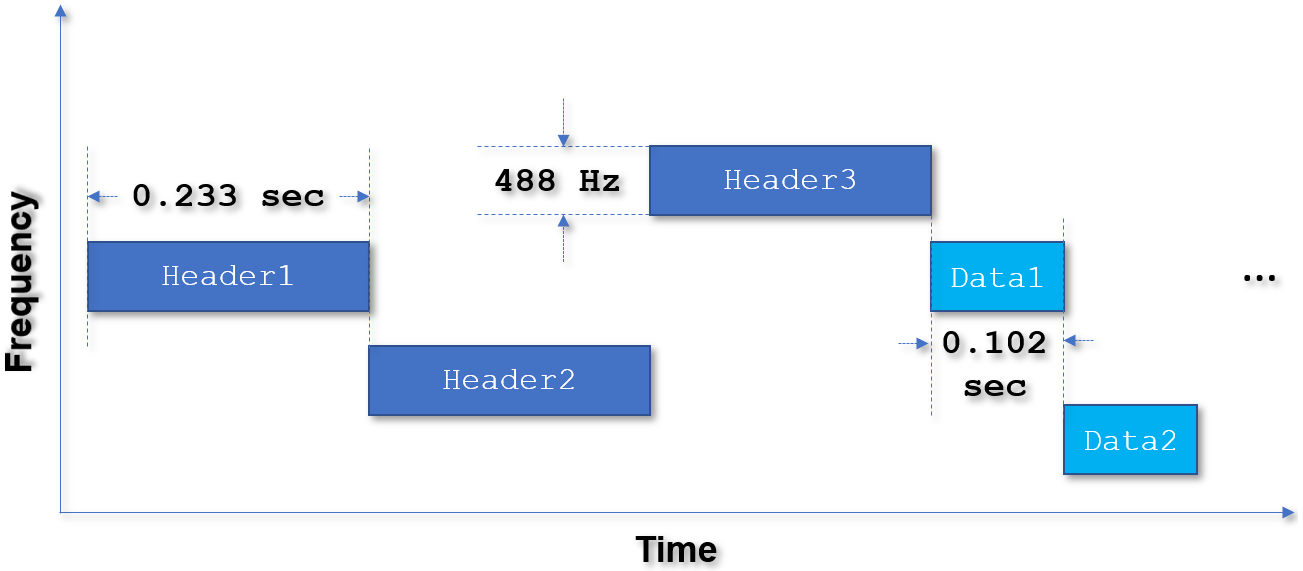}
	\end{center}
	\vspace{\figspacesub}
	\caption{A packet with DR8.}
	\label{fig:illus_pkt}
\end{figure}

In LR-FHSS, a node transmits signals with a very narrow bandwidth of 488 Hz, which is referred to as Occupied Band Width (OBW). The modulation is Gaussian Minimum Shift Keying (GMSK), where each symbol modulates 1 bit. The symbol duration, denoted as $\smbltime$, is about 2 ms. Bit `0' is represented by a linear phase decrease of  $\pi/2$ during the symbol time and `1' a linear phase increase of $\pi/2$. The phase change is smoothed by a Gaussian filter, so that the amount of phase change could be less than $\pi/2$ if two consecutive bits are different. Fig.~\ref{fig:ex_GMSK} shows an example of 6 symbols found in a packet trace, where the center of each symbol has been indicated with a marker and the symbol boundaries roughly align with the vertical lines.

A packet in LR-FHSS consists of headers and data fragments transmitted back to back on different frequencies, as shown in Fig.~\ref{fig:illus_pkt}. The header is always 0.233 seconds. For robustness, the header is repeated multiple times, which are 3 and 2 for DR8 and DR9, respectively, so that the gateway can detect the packet as long as one of the headers is received correctly. The data fragments are are not repeated. There could be multiple fragments depending on the size of the payload, where each fragment is always 0.102 seconds, except the last which may be shorter because it may carry less bits.  

\begin{figure} 
	\begin{center}
		\includegraphics[width=3in]{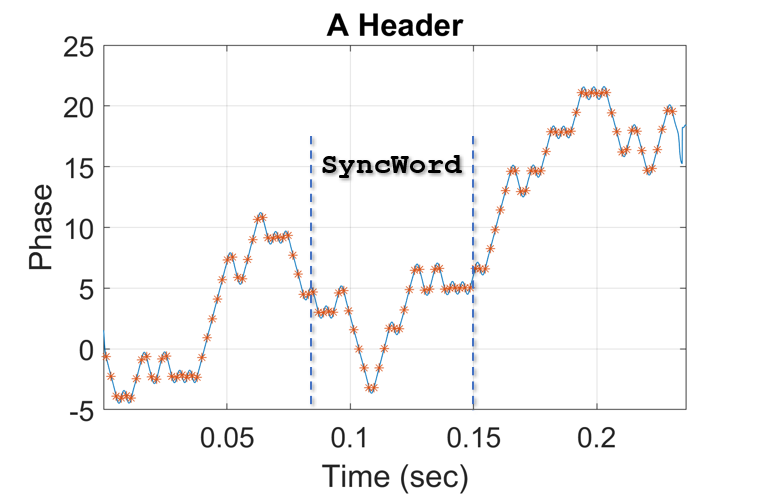} \\
		 {\small (a)}\\
		\includegraphics[width=3in]{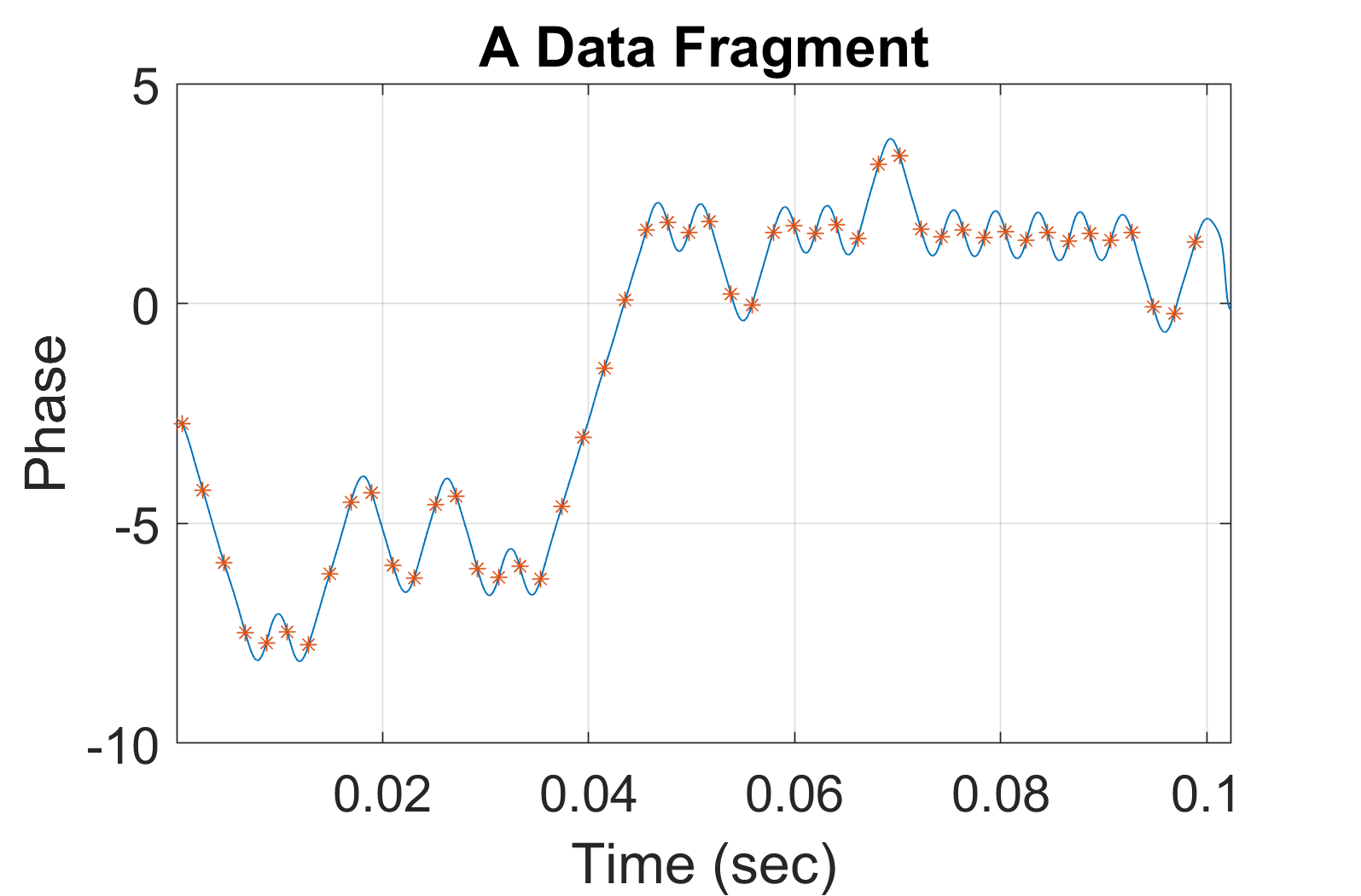} \\
		{\small (b)}
	\end{center}
	\vspace{\figspacesub}
	\caption{The phase signals. (a). Header. (b). Data fragment.}
	\vspace{-0.1in}
	\label{fig:ex_header}
\end{figure}

A demodulated header and data fragment are shown in Fig.~\ref{fig:ex_header}. The header carries key information, such as the Coding Rate (CR), payload length, and a 9-bit hop sequence id, as well as an 8-bit CRC. With the hop sequence ID, the frequencies used by all headers and data fragments can be determined. The information and CRC are 40 bits in total, which are encoded into 80 bits by a convolutional code with rate 1/2 and memory depth 4. The 80-bit codeword is then interleaved. A constant binary vector of length 32, called the {\fontfamily{qcr}\selectfont SyncWord}, is then inserted in the middle of the codeword, which allows the receiver to estimate the symbol boundaries and frequency. The payload bits are first added with a 16-bit CRC, padded with 6 `0's, then encoded by a convolutional code with memory depth 6, where the rates are 1/3 and 2/3 for DR8 and DR9, respectively. The codeword is then whitened, interleaved, and split into data fragments. 

\begin{figure} 
	\begin{center}
		\includegraphics[width=3.5in]{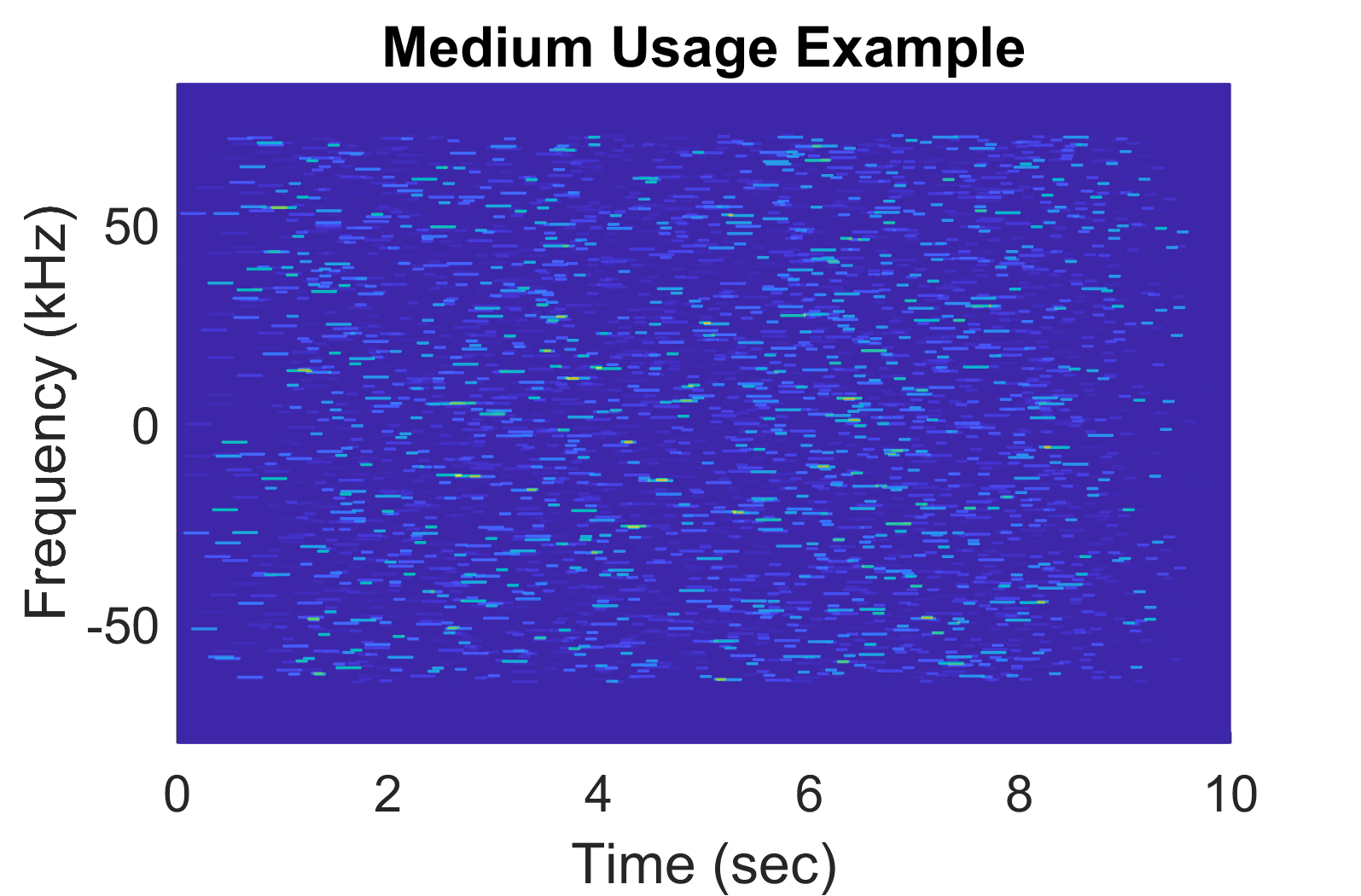} 
	\end{center}
	\vspace{\figspacesub}
	\caption{The medium usage map of a synthesized trace.}
	\vspace{-0.1in}
	\label{fig:ex_mediumusage}
\end{figure}

The system bandwidth supports 280 {\em channels}, where each channel is 488 Hz. Channels $k$, $k+8$, $k+16$, ..., $k+272$ are in {\em group} $k$ where $1 \le  k \le 8$. To transmit a packet, a node randomly selects a group and hops only among channels belonging to the same group. Fig.~\ref{fig:ex_mediumusage} shows the medium usage of a synthesized trace containing 400 DR8 packets with payload size randomly distributed between 8 and 16 bytes. It can be seen that the medium is fairly busy in this trace;  still, our receiver was able to decode over 90\% of the packets.   

LR-FHSS is very simple for the node, because a node only needs to encode the data and randomly select the transmission frequencies. The Medium Access Control (MAC) layer is basically ALOHA, i.e., a node can transmit a packet at any time of its choice. There is no need to maintain tight timing or frequency synchronization with the gateway, because packets can be transmitted at any time and frequency errors are equivalent to moving the signals to different frequency channels. Multiple access is supported by the physical layer due to the narrowness of the channel and random choices of frequency, which result in a low level of collision when the traffic load is not high. Long communication range is achieved also due to the narrowness of the channel because the noise power is proportional to the communication bandwidth.  

\section{Packet Trace Collection}
\label{sec:tracecollect}

\begin{figure}
	\centering
	\includegraphics[width=0.9in]{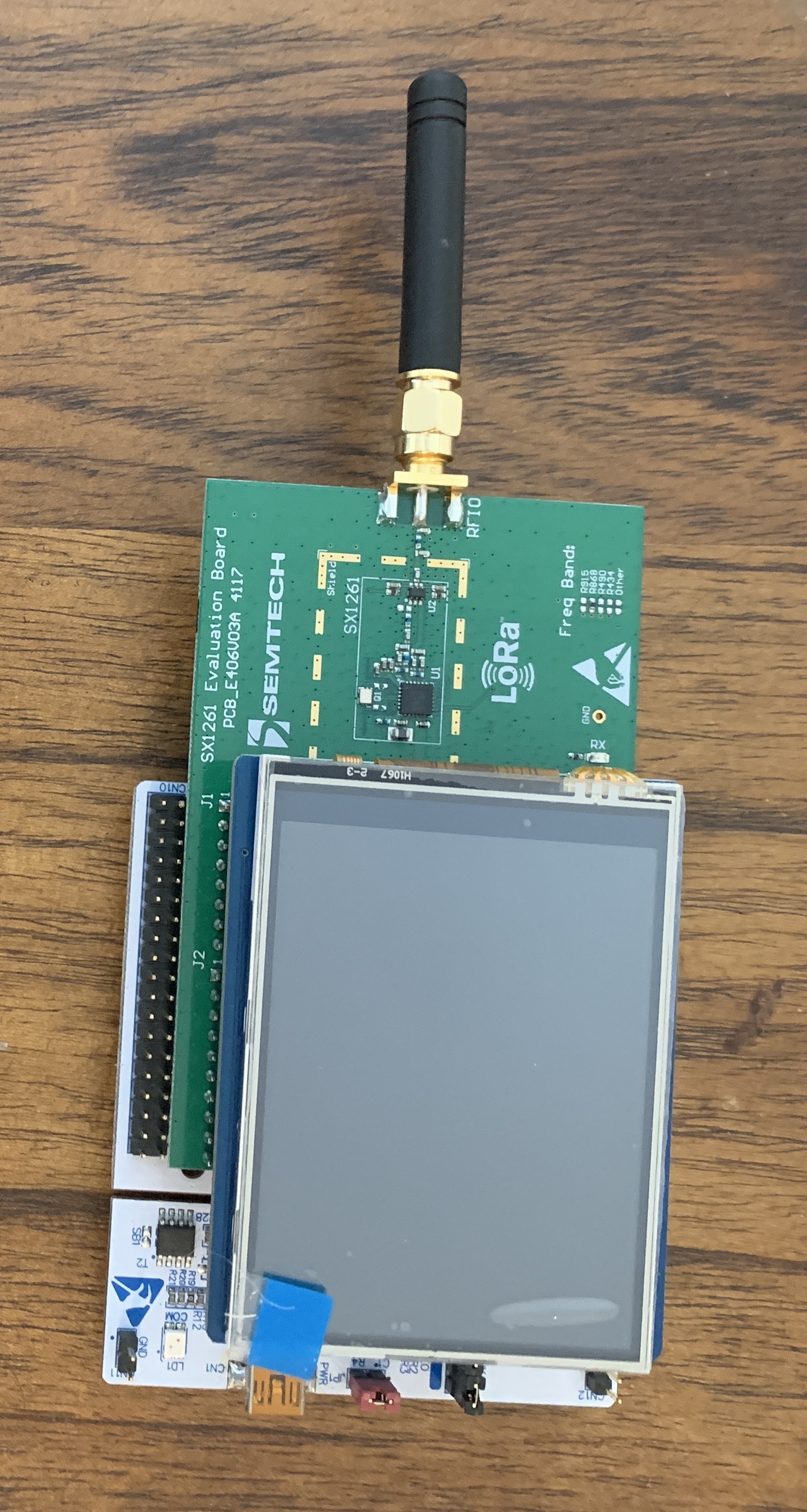}
	\hspace{0.1in}
	\includegraphics[width=2.25in]{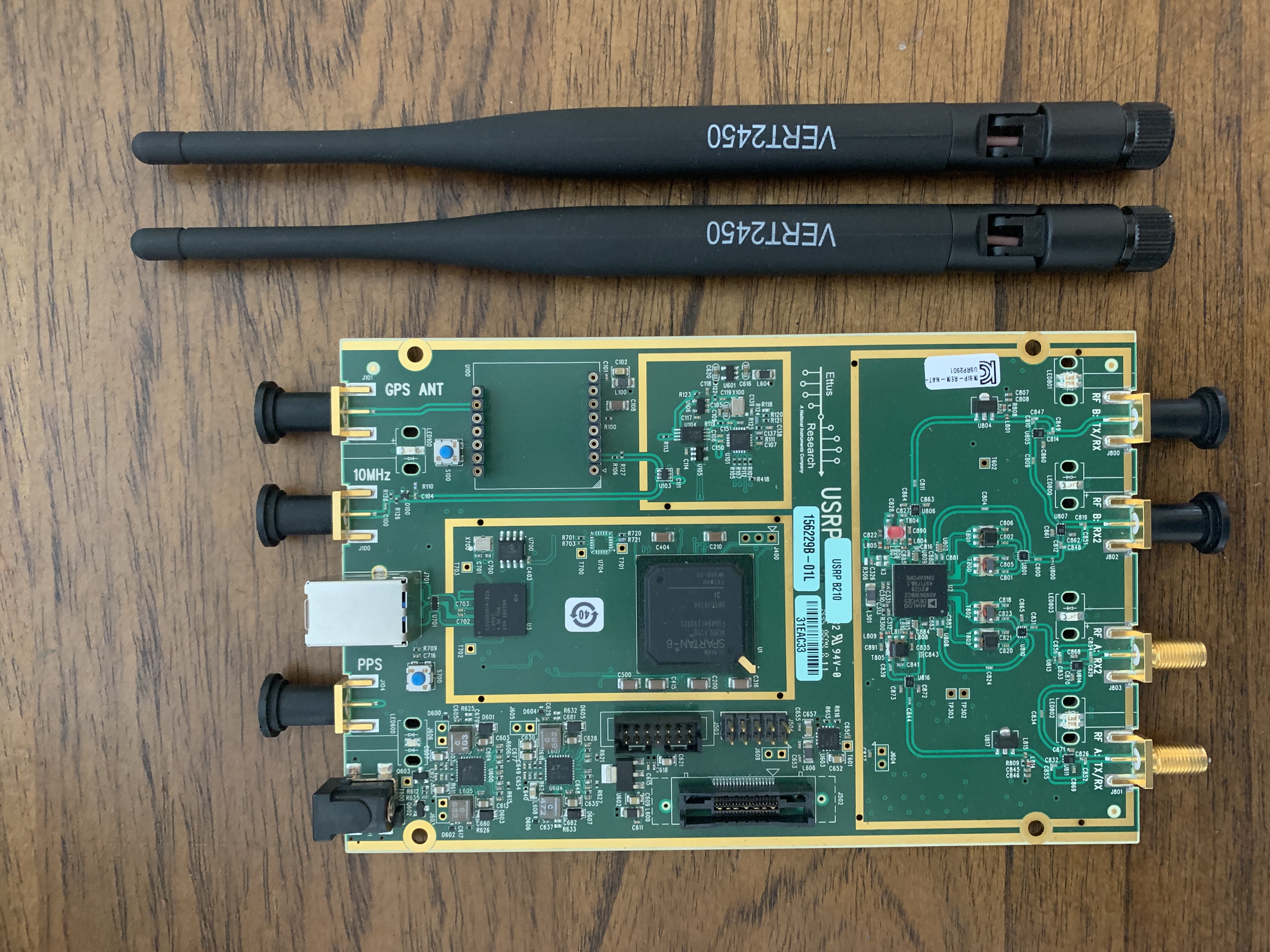} \\
	\hspace{-0.6in}  {\small (a)} \hspace{1.5in} {\small (b)}
	\vspace{-0.1in}
	\caption{(a). SX1261. (b). USRP B210.}
	\label{fig:devices}
\end{figure}

\begin{figure}
	\centering
	\includegraphics[width=3in]{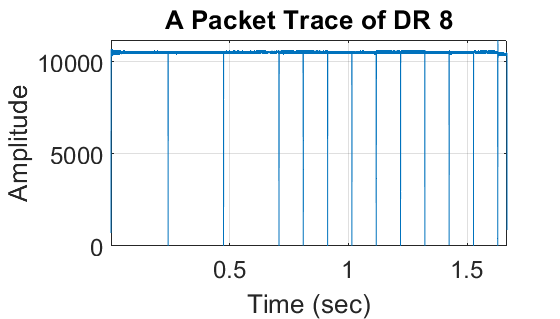} \\
	\vspace{-0.1in}
	\caption{A packet trace with DR8.}
	\label{fig:ex_trace}
\end{figure}

We collected packet traces with SX1261~\cite{LRFHSSdevice} as the transmitter and a USRP B210~\cite{USRPB210} as the receiver. Fig.~\ref{fig:devices}(a) and (b) show the SX1261 and USRP, respectively. 

We placed the transmitter and receiver close to each other with a line-of-sight path, so that the signal from SX1261 was very strong and could be used to approximate clean signals from an LR-FHSS transmitter. Clean signals are preferred because, as mentioned earlier, they allow radios in the POWDER platform~\cite{RefPowder} to emulate LR-FHSS nodes, as well as facilitating trace-driven simulations. To elaborate, as the traffic is random, in a synthesized trace, the number of overlapping packets could fluctuate largely over time, e.g., between 20 and 60. As the signal as well as noise in all traces are added, if the packet traces are noisy, the actual SNR in synthesized trace could also fluctuate, making it difficult to run an evaluation under a given SNR.  In addition, the traces may be processed by different channel models. If the traces were collected in a strong multi-path environment, the existing multi-path in the trace may interfere with the channel model. The trace of a packet with DR8 is shown in Fig.~\ref{fig:ex_trace}, which has 3 header replicas and 10 data fragments. It can be seen that the signal was very clean and stable. There exist natural separations between the headers and data fragments because LR-FHSS inserts a small idle period between consecutive headers and data fragments. 

For both DR8 and DR9, 500 packets were collected. The payload length of a packet was between 8 to 16 bytes, which does not include the 16-bit CRC. The number of packets of each size was either 55 or 56 so that the total number of packets of each data rate adds up to 500. Depending on the payload length, the number of data fragments with DR8 and DR9 were between 6 to 10 and 3 to 5, respectively. The transmission times of a packet with DR8 and DR9 were between 1.26 sec to 1.67 sec and 0.75 sec to 0.95 sec, respectively. The content of the payload and the hop sequence ID were randomly generated for each packet. Packet generation was reasonably convenient with SX1261, because SX1261 uses its driver to prepare the packet in software. As we had access to the driver code, in particular, {\fontfamily{qcr}\selectfont sx126x\_lr\_fhss\_ping.c}, we could freely modify the content as well as system parameters of the packet, such as transmission power, carrier frequency, code rate, bandwidth, number of headers, etc.  

During the trace collection, the system bandwidth was 137 kHz, the carrier frequency was 915 MHz, and the sampling rate was 500 ksps. For each raw trace, the start and end of the packets were found based on energy level and only signals containing the packets were kept. Each packet was then written to a trace file to be stored. As each sample is 4 bytes with 2 bytes for both the real and imaginary parts, a trace file is between 1.5 MB to 3.5 MB.

\section{Receiver Design}
\label{sec:receiver}

We designed and implemented a software LR-FHSS receiver from scratch, with which the performance of LR-FHSS can be evaluated. The receiver design is explained in this section.

\subsection{Overview} 

\begin{figure} 
	\begin{center}
		\includegraphics[width=5in]{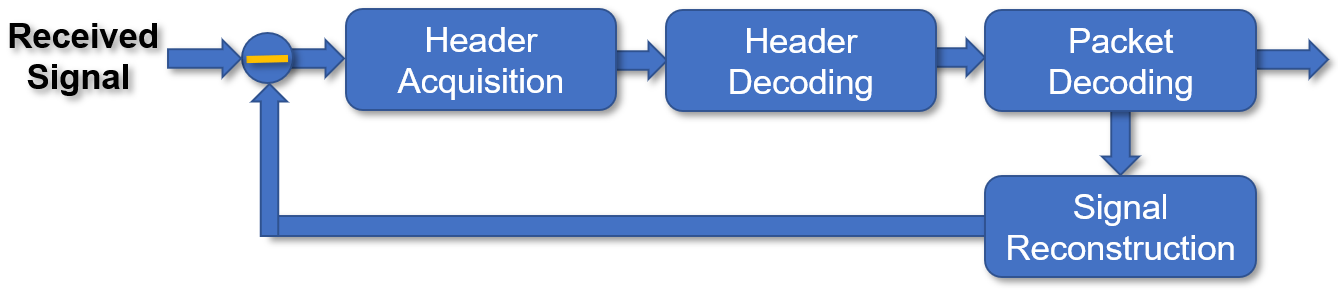} 
	\end{center}
	\vspace{\figspacesub}
	\caption{The receiver structure.}
	\vspace{-0.1in}
	\label{fig:receiver}
\end{figure}

A high level structural view of our receiver is shown in Fig.~\ref{fig:receiver}. The time-domain signal is first processed by the header acquisition component, which detects headers in the signal and finds the symbol boundaries and frequencies. The detected headers are then demodulated by the header decoding component. After this point, multiple headers belonging to the same packet are consolidated as one detected packet. Then, the detected packets are decoded by the packet decoding component. After a packet has been decoded, Successive Interference Cancellation (SIC) is performed, i.e., the signal of the packet is reconstructed by the signal reconstruction component and subtracted from the received signal so that more packets can be detected and decoded.

\subsection{Challenges and Contributions} 

When we started working on LR-FHSS, the complete information of LR-FHSS was not available. We got most help from a paper~\cite{9422331}, a document by Semtech~\cite{LoRaPara}, and the driver code of SX1261~\cite{sx126xdriver}. However, the information was limited to the sender side at a high level, such as the error correction code, interleaving, whitening, etc., while no information could be found about the receiver on issues such as packet detection and signal demodulation. 

We overcame the lack of documentation with a bit of reverse engineering and luck. One such example is related to {\fontfamily{qcr}\selectfont SyncWord}, which is needed to estimate the symbol boundary and frequency of the packet. We learned in~\cite{LoRaPara} that the value of {\fontfamily{qcr}\selectfont SyncWord} is {\fontfamily{qcr}\selectfont0x2C0F7995}. We also knew that {\fontfamily{qcr}\selectfont SyncWord} is in the header. However, at the time, the exact location of {\fontfamily{qcr}\selectfont SyncWord} was unclear. Nevertheless, we started playing with the trace signal. It was fairly easy to identify the start of the header, which was a sharp rise of energy. Then, we manually tried different frequency offsets, because we knew that when the correct frequency offset is applied, the phase signal should resemble those shown in Fig.~\ref{fig:ex_header}. Once that happened, we happily found that {\fontfamily{qcr}\selectfont SyncWord} matches the waveform in the middle of the header, which was a surprise to us because we thought it was at the beginning of the header like preambles in most other wireless networks. In fact, in the own document of Semtech~\cite{LoRaPara}, {\fontfamily{qcr}\selectfont SyncWord} was shown at the beginning of the header structure in a figure, which has likely propagated to other documents and papers such as~\cite{9422331}. 


As mentioned earlier, the challenges also include designing the receiver to achieve good performance, as there was no existing literature on the receiver design for LR-FHSS. Therefore, our contribution also include designing and implementing a functional receiver, as well as proposing novel solutions to improve the performance of LR-FHSS. To elaborate, some of the problems could be solved by standard library functions. For example, we were able to use the Matlab {\fontfamily{qcr}\selectfont lowpass} and {\fontfamily{qcr}\selectfont vitdec} functions to filter the signal and decode the convolutional code of the data, respectively. However, most of the problems could not be solved in this manner. For example, header acquisition requires a customized solution because it is specific to LR-FHSS. Some of other problems, such as demodulation and SIC, do not have good matching library functions, for which we had to design our own solutions. We also proposed a novel method, called Collision-Aware Erasure Decoding (CAED), which improves the error correction performance with little additional complexity.

\subsection{Header Acquisition} 

Header acquisition refers to detecting a header and finding key parameters about the packet, including symbol boundary and frequency. Header acquisition is challenging because the header could start at any time on any frequency, and there could be many headers and data fragments transmitted at the same time. 

\subsubsection{Initial Screening} 

\begin{figure} 
	\begin{center}
		\includegraphics[width=4in]{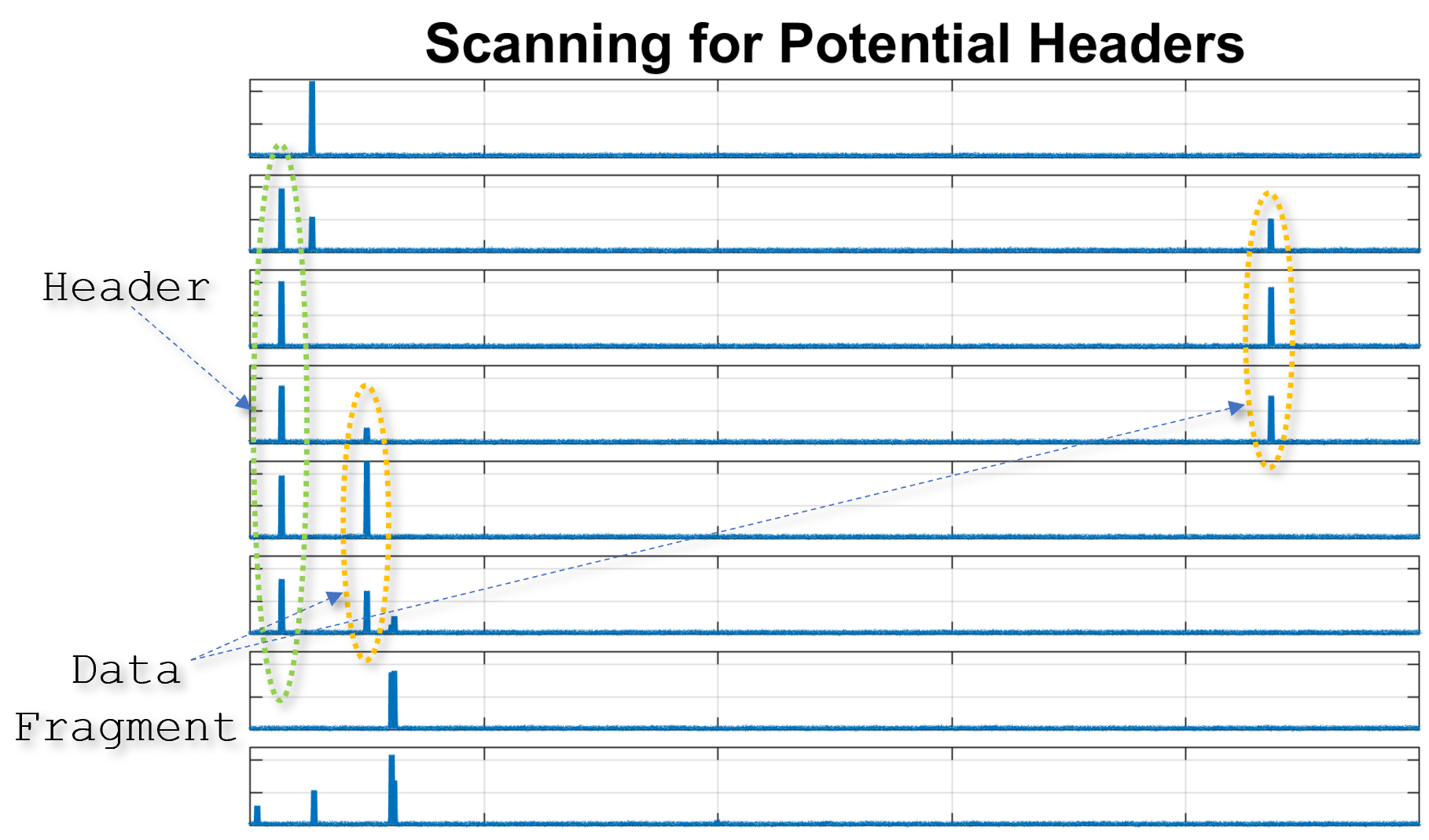} 
	\end{center}
	\vspace{\figspacesub}
	\caption{The initial screening of headers.}
	\vspace{-0.1in}
	\label{fig:ex_headerscan}
\end{figure}

Our method starts with an initial screening step that spots potential {\em candidate headers}. The idea is to scan for high peaks in the frequency domain, because LR-FHSS signal is in a narrow band with energy focused on a small number of consecutive FFT coefficients. The challenge is to separate the header from the data fragments, because both produce such high peaks. To solve this problem, the time-domain signal is divided into segments of $\screenseglen$ seconds where $\screenseglen=0.05$, and the FFT is performed on each segment. A header most likely produces peaks at the same or very close locations in 5 or more consecutive segments, because a header is 0.233 seconds. A data fragment, on the other hand, produces peaks in at most 3 consecutive segments, because a data fragment is 0.102 seconds. Therefore, a candidate header is identified if a {\em header pattern} is spotted, where header pattern is defined as peaks at very close locations in 5 or more consecutive segments. For example, Fig.~\ref{fig:ex_headerscan} shows the FFTs of 8 consecutive segments, in which a header pattern has been marked. Two data fragments have also been marked, which belong to the same packet. 

When a header pattern is spotted, a candidate header is added to a list. The initial estimated start time of the header is simply the start time of first segment where the header pattern is found. The initial estimated frequency can be calculated based on the location of the peak. The maximum timing error at this point is no more than $\screenseglen$ seconds, which may occur when the header starts at the end of a segment but is still identified because the signal is very strong. It should be noted that the frequency domain signal is always within the channel bandwidth but may exhibit peaks at random locations. A Gaussian filter is therefore used to smooth the frequency domain signal, which merges the energy within the signal bandwidth into one high peak at the center, so that the frequency error is typically within 100 Hz.     

\subsubsection{Coarse Timing Estimation} 

\begin{figure} 
	\begin{center}
		\includegraphics[width=3.5in]{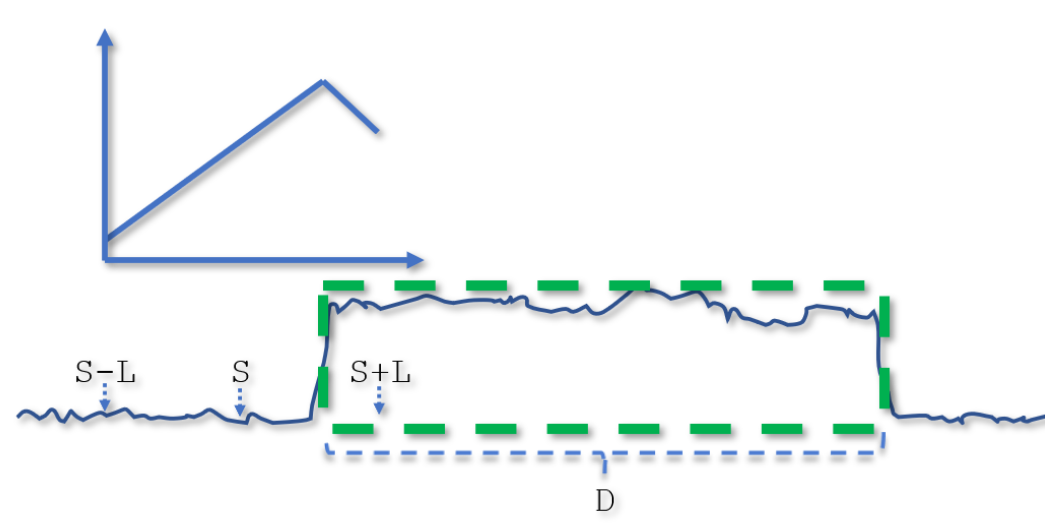} \\
		{\small (a)} \\
		\includegraphics[width=3.5in]{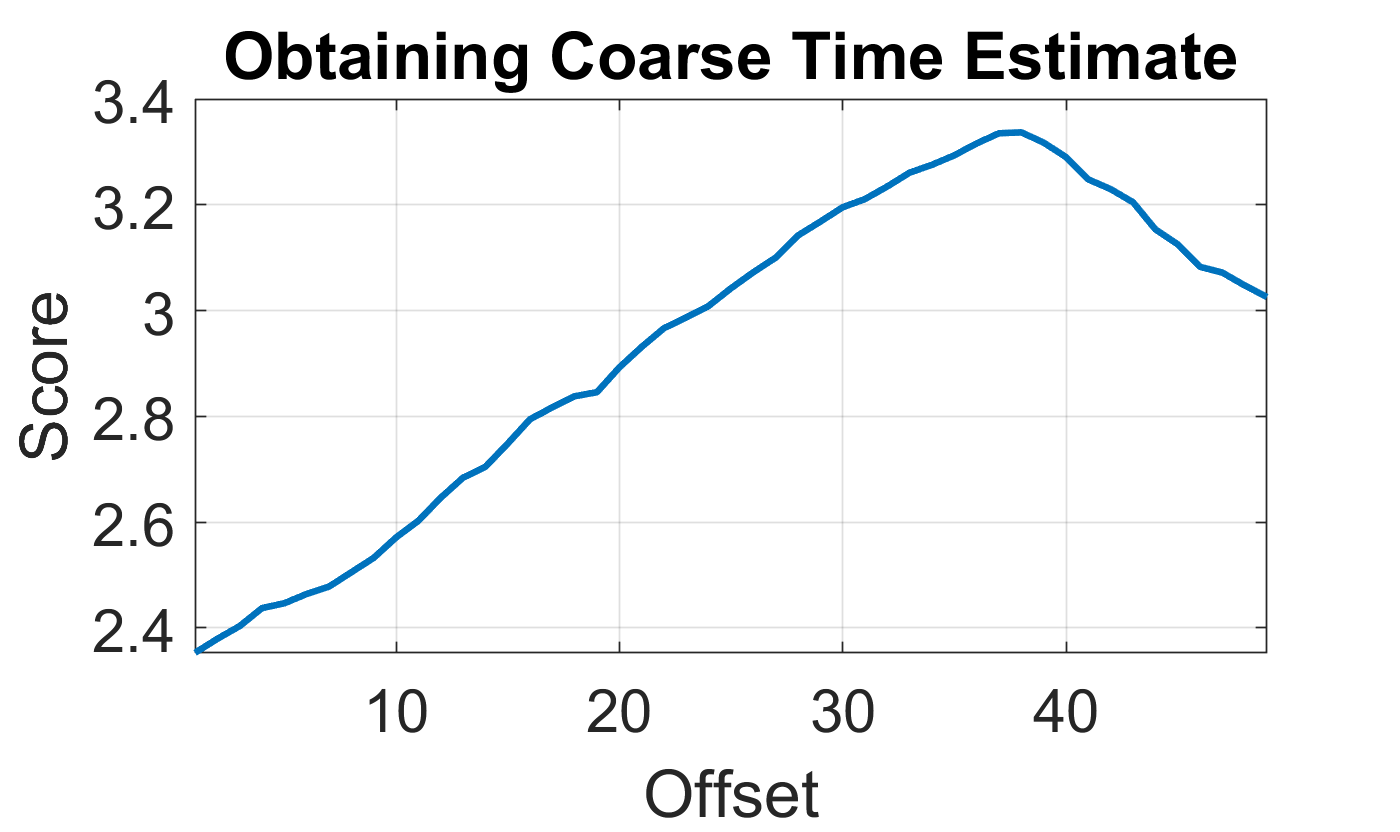} \\
		{\small (b)}
	\end{center}
	\vspace{\figspacesub}
	\caption{(a). Sliding window during coarse timing estimation. (b). The score as a function of the offset.}
	\vspace{-0,1in}
	\label{fig:ex_coarsetimeest}
\end{figure}

The next step obtains coarse estimates of the start times of the candidate headers. Suppose a candidate header has been spotted with an initial estimated start time of $\estt$. The idea is to slide a window of length $\hdrlen$ seconds, where $\hdrlen$ is the length of the header, starting from $\estt-\screenseglen$ and stopping at $\estt+\screenseglen$. As shown in Fig.~\ref{fig:ex_coarsetimeest}(a), when the window overlaps with the header exactly, maximum signal energy from the header is observed. Therefore, for a window that starts at $\estt-\screenseglen+x$, where $x$ is called the {\em offset}, the {\em score} is defined as the total amount of energy inside the window within the frequency channel of the header. To reduce the timing error to the level of symbols, the window slides down $\smbltime$ seconds at a time, where $\smbltime$ denotes the symbol time. Fig.~\ref{fig:ex_coarsetimeest}(b) shows the actual score when performing the scan on a header.

It should be mentioned that before the search is performed, the time-domain signal should be first down-converted then passed through a Low-Pass-Filer (LPF), so that the energy of the header could be examined in the baseband without the interference and noise on other frequencies. The down-conversion is simply to divide the time-domain signal by a sinusoid whose the frequency is the initial estimated frequency of the header, so that the header signal is moved to the baseband, i.e., within $\pm$244 Hz. The LPF then rejects signals outside the baseband, which includes signals from other packets, as well as most of the noise. Removing noise outside the baseband gives the signal a huge SNR boost, because the noise power is proportional to the signal bandwidth. That is, as the OCW is 137 kHz while the OBW is only 488 Hz, after the LPF, the SNR is increased by 280 times, which is roughly 25 dB. As we could not find specifications about the LPF in LR-FHSS, we used a relatively ``high-end'' LPF in Matlab, namely, the {\fontfamily{qcr}\selectfont lowpass} function, which yields better performance than other LPFs we have tried. To be more specific, we used an {\fontfamily{qcr}\selectfont iir} filter where the cutoff frequency was 200 Hz,
{\fontfamily{qcr}\selectfont StopbandAttenuation} was 100, and  {\fontfamily{qcr}\selectfont Steepness} was 0.9. The cutoff frequency is slightly lower than half of OBW because of the transition band of the filter. 

\subsubsection{Fine Estimation} 

\begin{figure} 
	\begin{center}
		\includegraphics[width=4in]{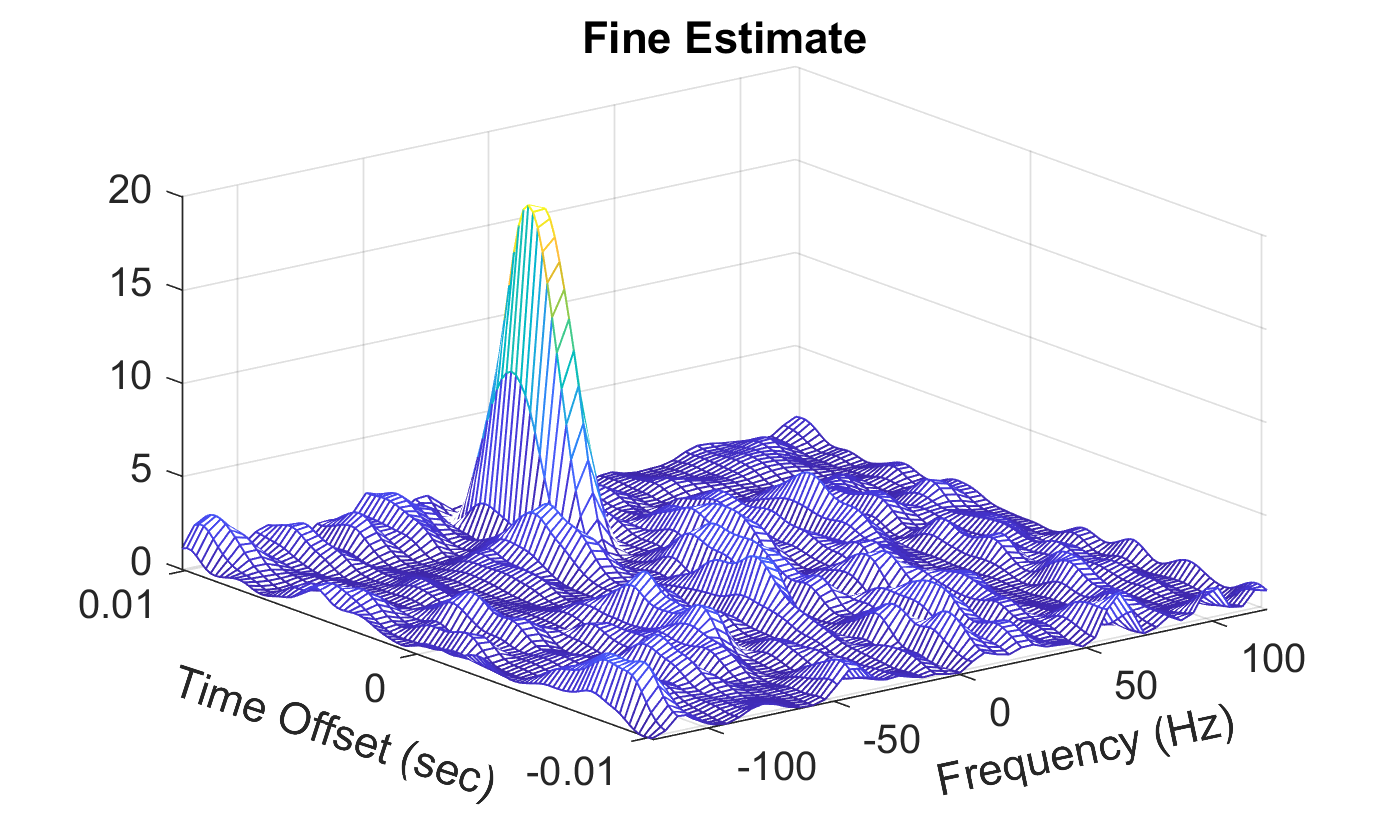} 
	\end{center}
	\vspace{\figspacesub}
	\caption{The inner product with {\fontfamily{qcr}\selectfont SyncWord} during fine timing and frequency estimation.}
	\vspace{-0.1in}
	\label{fig:ex_fineest}
\end{figure}

Lastly, for a candidate header, fine estimations of the symbol boundary and frequency are obtained. It is expected that, before this step, the timing and frequency errors have been reduced to within 5$\smbltime$ seconds and 100 Hz, respectively. The objective of this step is to further reduce the errors to a level that allows for correct demodulation of the symbols, which, with our current implementation, are $\smbltime/10$ seconds and 5 Hz, respectively.

Unlike the initial screening and coarse timing estimation, fine estimation cannot be achieved based only on energy observations and must rely on the {\fontfamily{qcr}\selectfont SyncWord}. If the receiver has obtained correct symbol boundary and frequency, it can take samples at the locations of the {\fontfamily{qcr}\selectfont SyncWord}, which should match {\fontfamily{qcr}\selectfont SyncWord}, i.e., produce a high {\em inner-product} with {\fontfamily{qcr}\selectfont SyncWord}. Therefore, different combinations of timing and frequency offsets are applied to the header signal and the one with the highest inner-product is used as the fine estimate. To be more specific, for any combination, first, 32 samples are taken with an interval of $\smbltime$ seconds according to the timing offset. Second, the frequency offset is applied to the 32 samples. Third, the inner product of the 32-sample vector and {\fontfamily{qcr}\selectfont SyncWord} is calculated and the squared norm of the inner product is used as the score. To achieve the desired level of accuracy, an exhaustive search at a step of $\smbltime/10$ seconds and 5 Hz in the time and frequency domains is performed, where the time and frequency offset values are within $\pm 5 \smbltime$ seconds and $\pm$100 Hz, respectively. The computation cost is still reasonable because {\fontfamily{qcr}\selectfont SyncWord} is only 32 bits. Fig.~\ref{fig:ex_fineest} shows a typical search, where the peak, which corresponds to the best match, can clearly be seen.

\subsection{Demodulation} 
\label{sec:demod}

\newcommand{\samp}{\ensuremath{m}}

The demodulation step converts the waveform of each symbol into a real number to be used by the error correction decoder. The information can be extracted form the slope of the phase signal, because a negative phase change should indicate `0' and a positive change `1'. For simplicity, for each symbol, only two samples, denoted as $\samp_1$ and $\samp_2$, are used, which are to the left and right of the center of the symbol by $\smbltime/4$ seconds, respectively. The phase change from $\samp_1$ to $\samp_2$ is used as the estimate of the slope. Phase values near the boundaries of the symbol should be avoided because of the smoothing near the symbol boundary. It was found that using more than two samples only marginally improve the performance because the signal has already been through the low pass filter and over-sampling does not further improve the SNR. As the error correction decoder needs a soft demodulation value, the output is the {\em normalized slope}, which is the estimated slope divided by the ideal slope when the bit is `1'. For receiver with multiple antennas, the slope is estimated for each antenna individually then combined according to Maximum Ratio Combining (MRC), where the weight of an antenna is proportional to its channel strength.

\subsection{Error Correction with Collision-Aware Erasure Decoding (CAED)} 

LR-FHSS uses different convolutional codes for the header and the data fragments. The details of the encoders can be found in the driver code of SX1261~\cite{sx126xdriver}. In both cases, Viterbi decoders with soft decoding can be used. The encoder of the data always starts with initial state 0 and always pads `0's to the end of the data so that the encoder returns to state 0 at the end. Therefore, we were able to use the standard {\fontfamily{qcr}\selectfont vitdec} function in Matlab to decode the data. The encoder of the header, on the other hand, is different, for which we had to write our own code. To elaborate, recall that the memory depth of the header code is 4. The encoder may start with any of the 16 states as the initial state and does not pad `0's to the end. To decode the header, all 16 possible initial states are attempted. The one leading to the minimum cost is the estimated initial state, which will be used for signal reconstruction.

We further augment the decoder with a technique specifically designed for LR-FHSS, called Collision-Aware Erasure Decoding (CAED). Often, a packet cannot be decoded correctly because parts of it collides with another packet on a close frequency. Fortunately,  most of the collisions are known even before packet decoding, because as long as one of the headers of a packet is decoded correctly, its start time, end time, and frequency of each header and data fragment are known. Therefore, if most packets have been detected correctly, the entire time-frequency occupancy map such as that in Fig.~\ref{fig:ex_mediumusage} can be obtained, with which the collision conditions of every packet can be obtained to assist data decoding. Our idea is to mark any symbol under significant collision as an {\em erasure}, because such symbols can be toxic. That is, the soft value of the symbol is set to 0, so that it does not misguide the decoder. A symbol is considered under collision if the amount of energy observed on a frequency no more than 244 Hz away is more than 1.5 times of the energy of the packet to which this symbol belongs. For simplicity, the energy of a packet is simply the highest score found during the fine estimation step of header acquisition. The amount of energy on a frequency at a certain time is the total amount of energy of all packets that transmit on this frequency at the time. CAED is a very simple method that can bring noticeable gains, as illustrated in Section~\ref{sec:evaluation}. 

\subsection{Successive Interference Cancellation (SIC)} 

Successive Interference Cancellation (SIC) is a widely adopted method to reduce interference and improve network performance. The idea is that, after a packet has been decoded, its waveform can be regenerated and subtracted from the received signal, so that the interference it causes to other packets can be removed. Note that it is not difficult to regenerate the waveform to match the transmitted waveform, which is simply to pass the decoded packet through the modulation process. The challenge is to match the {\em received} waveform, which has been modified by the wireless channel and is subject to residual errors of symbol timing and frequency estimation. Such residual errors may not bother data demodulation but can lead to significant errors during signal reconstruction. 

We designed a customized SIC method for LR-FHSS that can be applied to both headers and data fragments. In the following, it is explained for data fragments. The input consists of the data bits as well as $\wave$, which is the received waveform of the fragment. $\wave$ should have been down-converted and passed through the LPF. Given the data bits in this fragment, the ideal waveform, denoted as $\wave^*$, can be obtained by linearly increasing or decreasing the phase according to the value of each bit. The core of the problem is to convert $\wave^*$ to match the signal in $\wave$ but not the noise.

\newcommand{\cham}{\ensuremath{a}}
\newcommand{\chph}{\ensuremath{\theta}}
\newcommand{\terr}{\ensuremath{\tau}}
\newcommand{\ferr}{\ensuremath{\delta}}

For simplicity, first, consider a single antenna system. The differences between $\wave$ and $\wave^*$ are mainly caused by three factors, namely, channel distortion, timing error, and frequency error. Channel distortion exists because the wireless channel always modifies the amplitude and phase of the signal. Fortunately, owing to the narrow bandwidth, the channel is flat, so that the channel state can be represented by a single complex number denoted as $\cham e^{i \chph}$, i.e., the received signal is basically the transmitted signal multiplied by $\cham e^{i \chph}$. As a reasonable estimate of $\cham$ is the magnitude of the received signal, the only parameter that needs to be estimated is the channel phase $\chph$. The timing error is denoted as $\terr$ and is defined as the difference between the actual start time of a symbol and the estimated start time of the same symbol. The residual frequency error is denoted as $\ferr$. As the data fragment has been decoded correctly, both $\terr$ and $\ferr$ are small.

\begin{figure} 
	\begin{center}
		\includegraphics[width=3.5in]{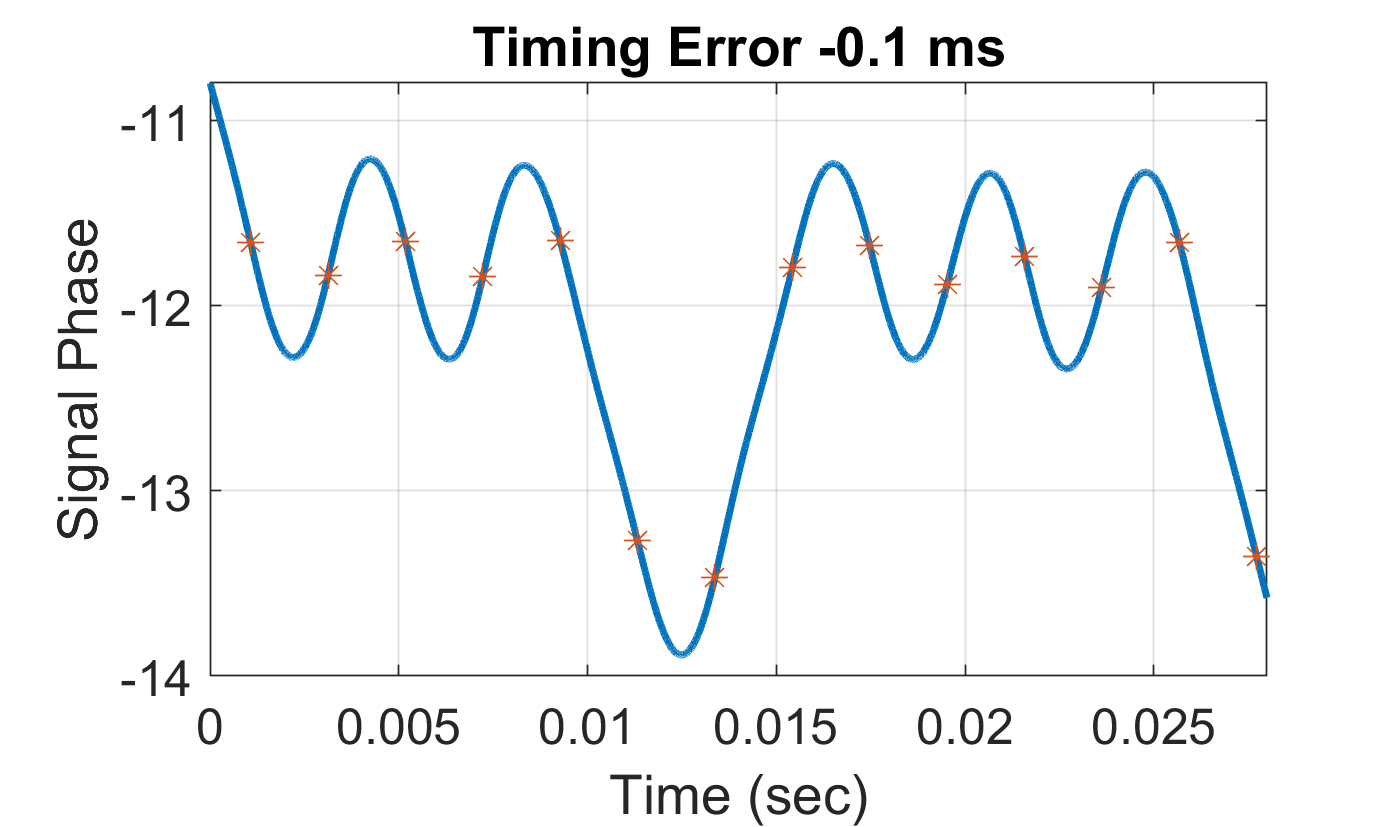} \\
		{\small (a)}\\
		\includegraphics[width=3.5in]{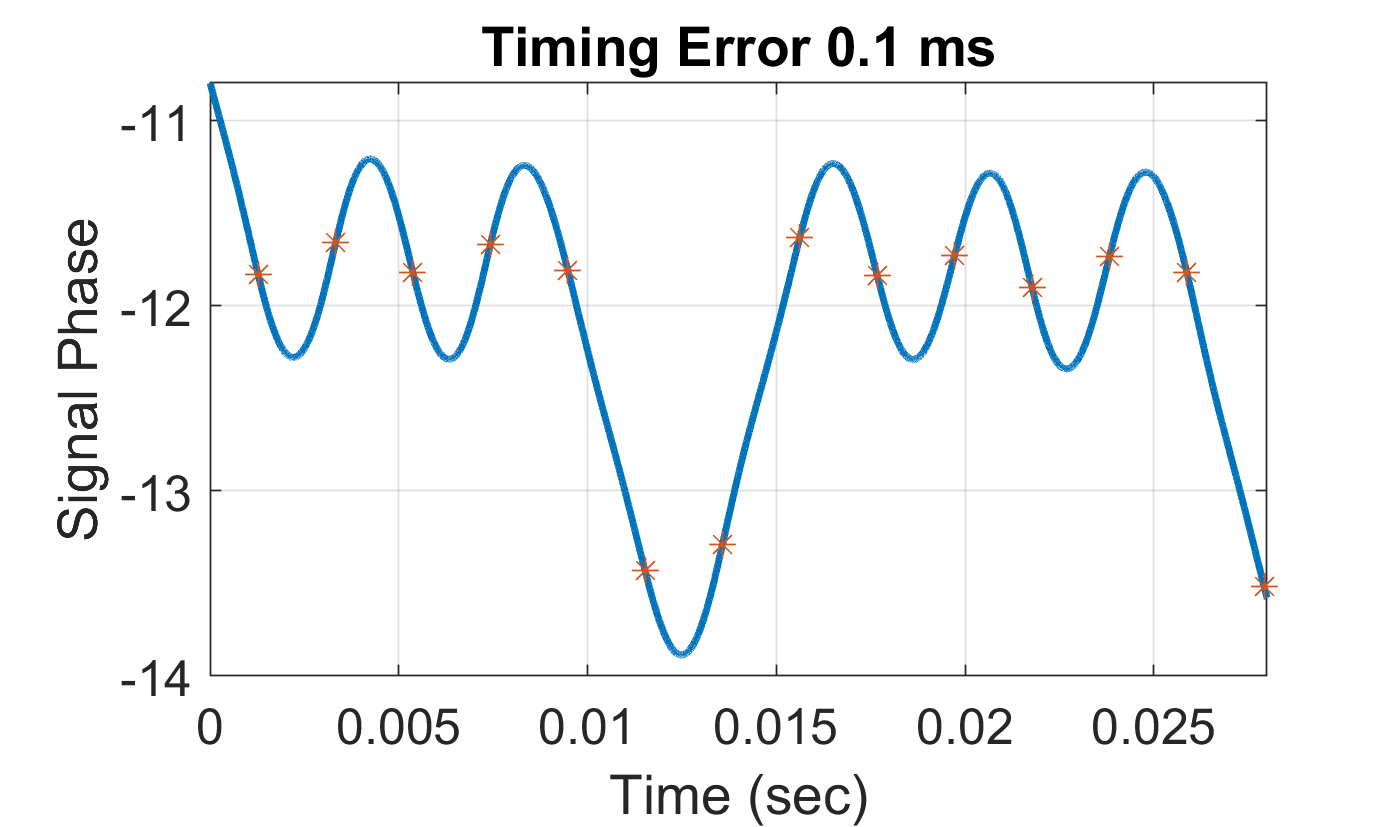} \\
		{\small (b)}
	\end{center}
	\vspace{\figspacesub}
	\caption{Phase error due to timing errors. (a). Negative error. (b). Positive error.}
	\vspace{-0.1in}
	\label{fig:ex_terr}
\end{figure}

Our method is a two-step process, where $\terr$ is estimated during Step 1, while $\ferr$ and $\chph$ are estimated during Step 2. We first explain Step 1, after which it should be clear why $\terr$ can be estimated without $\ferr$ and $\chph$. The estimation is based on transitions in the data from `0' to `1' and `1' to `0,' because the phase errors found in such transitions are proportional to $\terr$. To elaborate, define the phase of a symbol as the phase in the middle of the symbol. When $\terr=0$, the phases of two symbols in a transition should be the same, because the phase has been increased and decreased by the same amount. However, as shown in Fig.~\ref{fig:ex_terr}, if the estimated sample time is too early, i.e., $\terr < 0$, the phase difference is $-\pi \terr / \smbltime$ for a `0' to `1' transition and $\pi \terr / \smbltime$ for a `1' to `0' transition; similarly, if the estimated sample time is too late, i.e., $\terr > 0$, the amount of difference stays the same, but the signs are flipped. Therefore, based on the phase difference of the transitions, $\terr$ can be estimated. Note that $\chph$ is canceled during the phase difference calculation, while $\ferr$ does not lead to a significant bias because it is small. 

\begin{figure} 
	\begin{center}
		\includegraphics[width=3.5in]{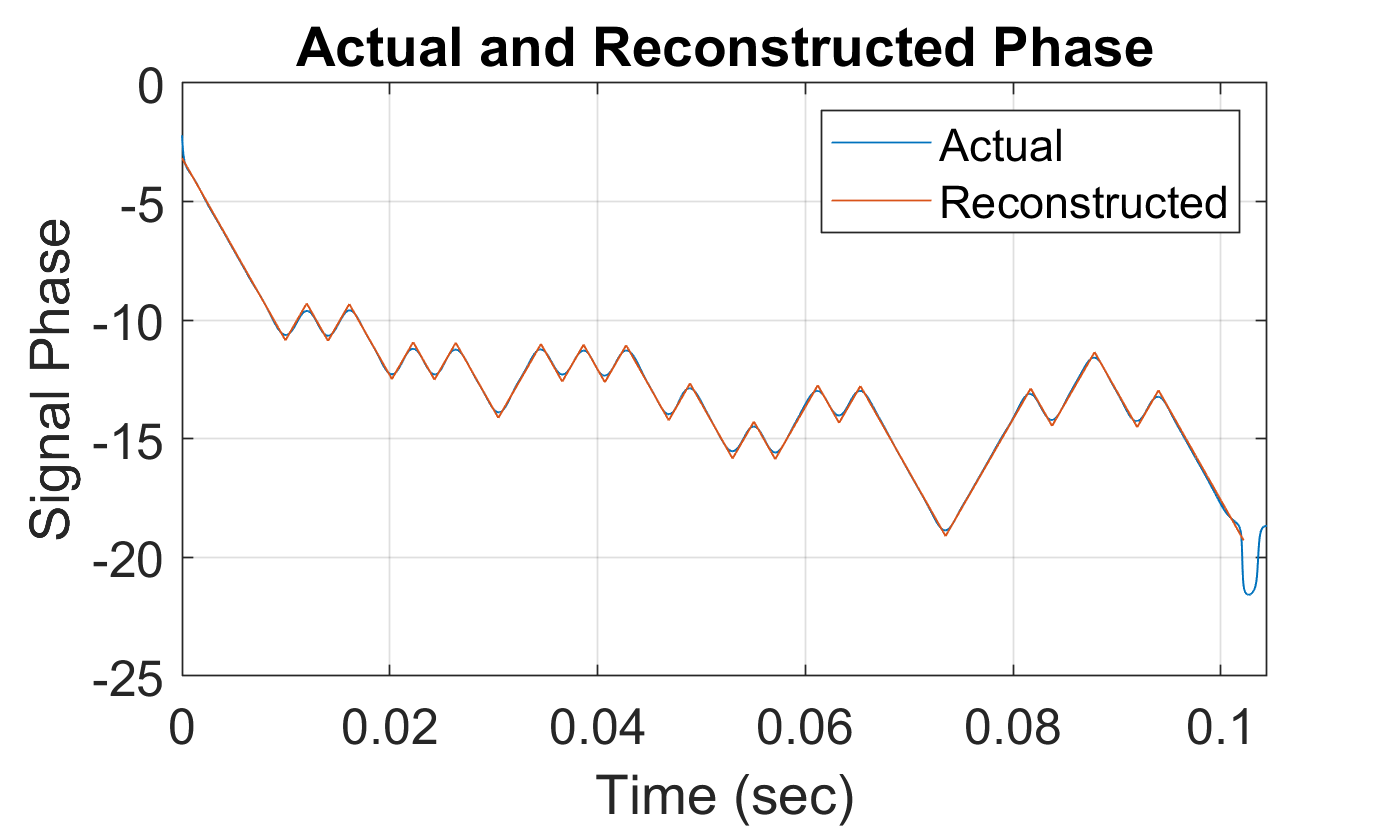} 
	\end{center}
	\vspace{\figspacesub}
	\caption{Actual and reconstructed phase signal of a data fragment.}
	\label{fig:ex_recon}
\end{figure}

In Step 2, first, the sample time is adjusted based on $\terr$. If there is no more timing error, $\wave = \wave^* \cham e^{i \chph} \odot \sinu_{\ferr} + n$, where $\sinu_{\ferr}$ denotes a sinusoid with frequency $\ferr$, $n$ denotes noise, and $\odot$ denotes element-wise multiplication of two vectors. Therefore, to estimate  $\ferr$ and $\chph$ is to find the phase difference of $\wave$ and $\wave^*$ and fit the difference with a line, where the slope and y-intercept are $\ferr$ and $\chph$, respectively. To accommodate phase jitters in certain channels, Step 2 is actually performed every 10 symbols, called a {\em recon-segment}. As an exmaple, Fig.~\ref{fig:ex_recon} shows a typical case, where the reconstructed signal matches the actual signal very well. 


\newcommand{\prcv}{\ensuremath{\Theta}}
\newcommand{\antn}{\ensuremath{A}}
\newcommand{\aidx}{\ensuremath{a}}
\newcommand{\tidx}{\ensuremath{t}}
\newcommand{\slope}{\ensuremath{k}}
\newcommand{\fitb}{\ensuremath{b}}
\newcommand{\sigl}{\ensuremath{N}}
\newcommand{\OBJ}{\ensuremath{G}}
\newcommand{\opw}{\ensuremath{w}}

Next, consider multiple antenna systems with $\antn$ antennas where $\antn>1$. $\terr$ can be estimated independently for each antenna and combined according to MRC. $\ferr$ should still be the same across antennas but the channel phase, denoted as $\chph_1$, $\chph_2$, $\ldots$, $\chph_\antn$, could be different. Therefore, they should be jointly estimated. To be more specific, let the number of samples used in the recon-segment be $\sigl$, which is currently 40. Let the difference of the received phase signal and the ideal phase be a matrix denoted as $\prcv$, i.e., $\prcv_{\aidx,\tidx}$ is the phase difference of the signal from antenna $\aidx$ at sample $\tidx$, where $1 \le \aidx \le \antn$ and $1 \le  \tidx \le \sigl$. As the signal strength of different antennas are different, a {\em weight} is assigned to each antenna depending on the signal strength, which is denoted as $\opw_\aidx$ for antenna $\aidx$ and normalized so that $\sum_{\aidx=1}^{\antn} \opw_\aidx = 1$. Mathematically, the problem can be formalized as finding $\ferr$  and $\chph_1$, $\chph_2$, $\ldots$, $\chph_\antn$ so that
\begin{equation}
	\OBJ = \sum_{\aidx=1}^{\antn} \sum_{\tidx=1}^{\sigl} \opw_\aidx (\prcv_{\aidx,\tidx} - \ferr \tidx -  \chph_\aidx)^2
\end{equation}
is minimized. Taking the partial derivative of $\OBJ$ with respect to $\chph_\aidx$ and set it 0, it can be found that
\begin{equation}
	\chph_\aidx = \frac{\sum_{\tidx=1}^{\sigl} \prcv_{\aidx,\tidx}}{\sigl} - \frac{(\sigl+1)\ferr }{2}.
\end{equation}
Similarly, taking the partial derivative of $\OBJ$ with respect to $\ferr$ and set it 0, it can be found that
\begin{equation}
	\ferr \sum_{\aidx=1}^{\antn}  \opw_\aidx \sum_{\tidx=1}^\sigl \tidx^2 = \sum_{\aidx=1}^{\antn} \sum_{\tidx=1}^{\sigl} \opw_\aidx \prcv_{\aidx,\tidx} \tidx  - \frac{\sigl(\sigl+1)}{2}\sum_{\aidx=1}^{\antn} \opw_\aidx \chph_\aidx, 
\end{equation}
which leads to
\begin{equation}
	\ferr  = \frac{\sum_{\aidx=1}^{\antn} \sum_{\tidx=1}^{\sigl} \opw_\aidx \prcv_{\aidx,\tidx} \tidx - (\sigl+1)/2\sum_{\aidx=1}^{\antn} \sum_{\tidx=1}^{\sigl} \opw_\aidx \prcv_{\aidx,\tidx}}{\sigl(\sigl+1)(2\sigl+1)/6 - \sigl(\sigl+1)^2/4}.
\end{equation}

 \section{Experimental Demonstration}
 \label{sec:powderexp}
 
  \begin{figure} 
 	\begin{center}
 		\includegraphics[width=3in]{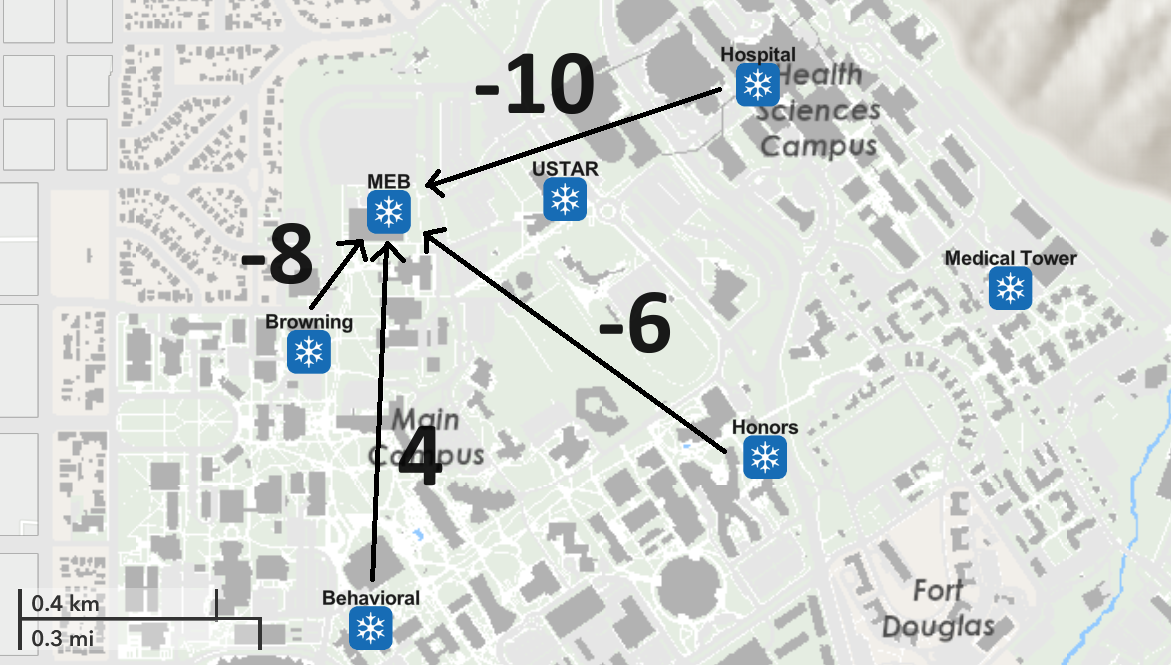} 
 	\end{center}
 	\vspace{\figspacesub}
 	\caption{Locations of the radios and link SNR.}
 	\vspace{-0.1in}
 	\label{fig:powderexpmap}
 \end{figure}
 
 \begin{figure} 
 	\begin{center}
 		\includegraphics[width=3.5in]{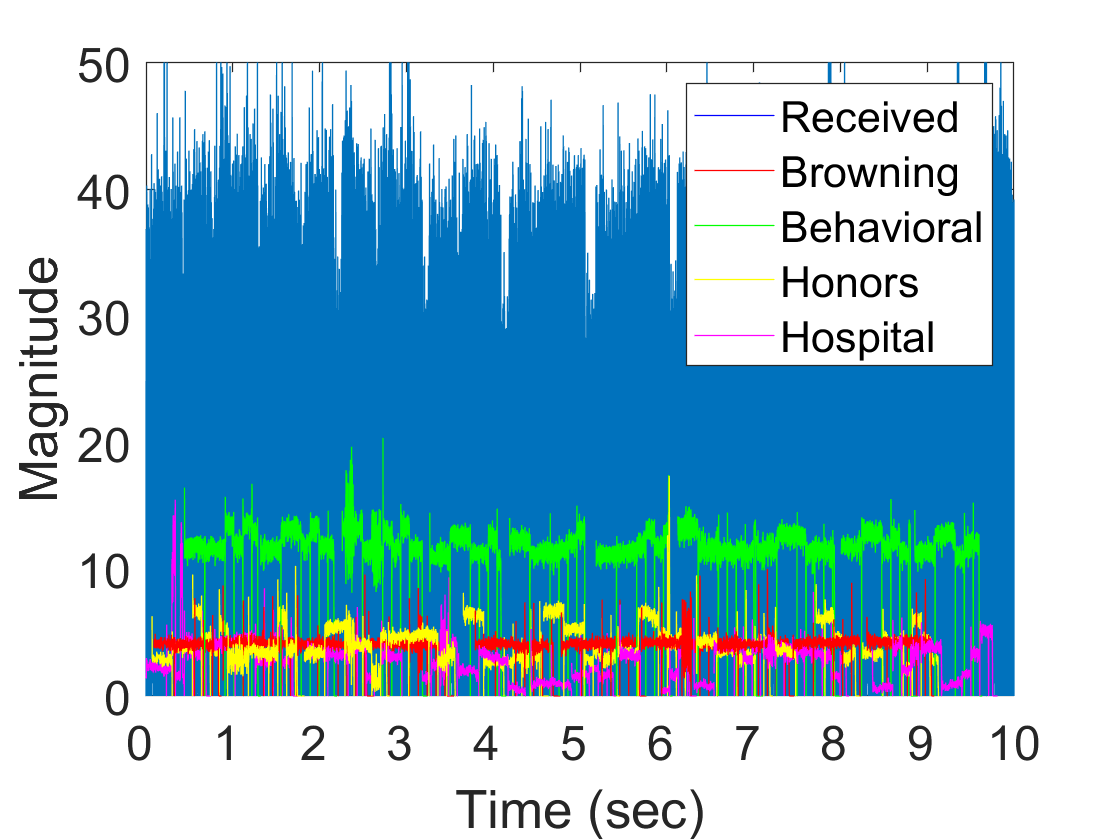} 
 	\end{center}
 	\vspace{\figspacesub}
 	\caption{The received signal and estimated signals from individual nodes.}
 	\label{fig:powderexpsig}
 \end{figure}

\begin{figure} 
	\begin{center}
		\includegraphics[width=3in]{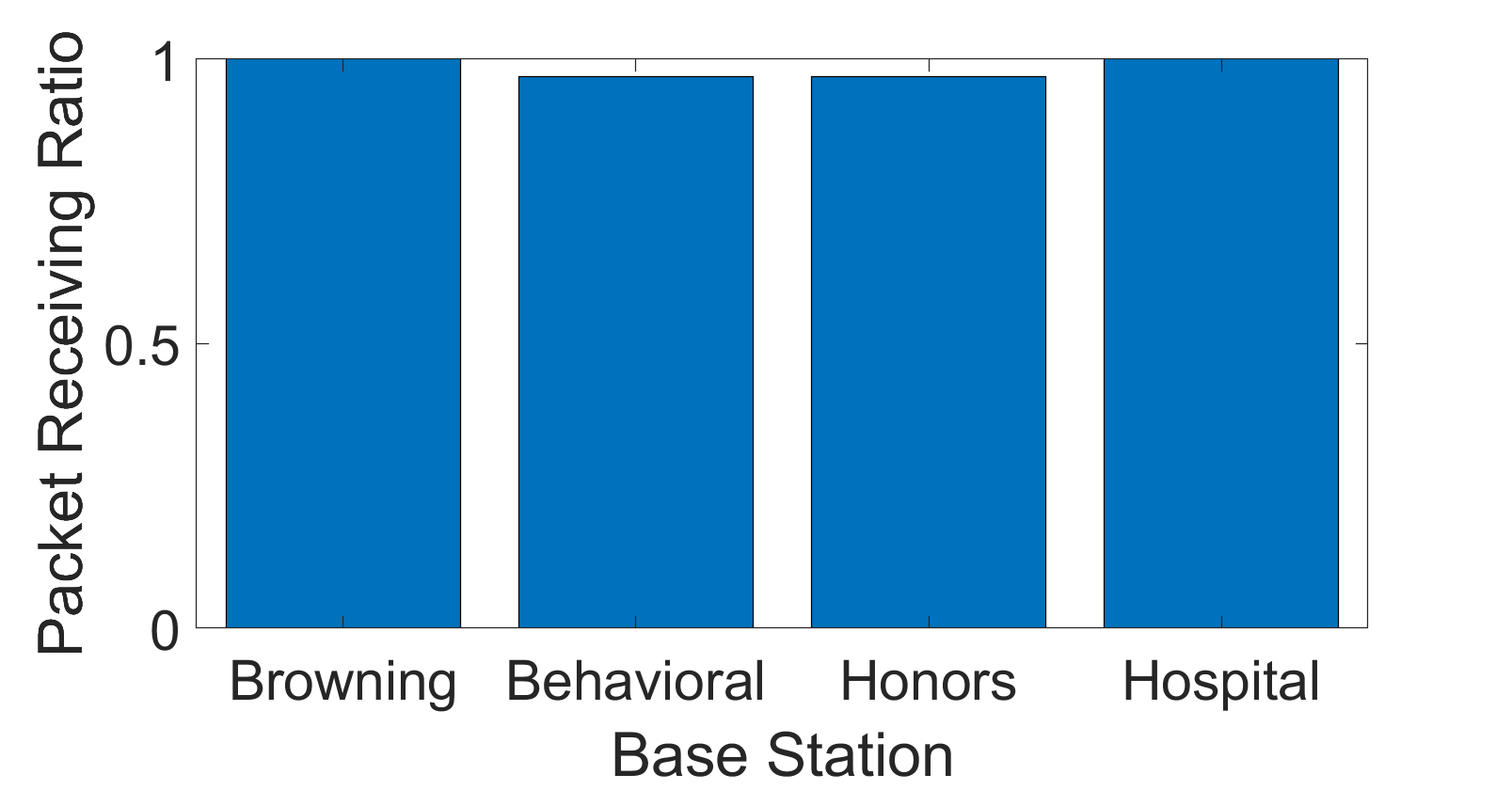} 
	\end{center}
	\vspace{\figspacesub}
	\caption{The PRRs of the nodes in the experiment.}
	\vspace{-0.1in}
	\label{fig:powderexpres}
\end{figure} 
 
We first demonstrate our receiver design and implementation with a real-world experiment in the POWDER wireless platform~\cite{RefPowder}. We set up 4 radios as nodes and one radio as the gateway. The radios in POWDER are USRPs and can transmit LR-FHSS packets by playing each packet trace as a file. Fig.~\ref{fig:powderexpmap} shows the locations of the radios and the estimated SNR of the links during the experiments measured in dB. The link SNRs were not proportional to the link distances in part because different transmitter gains were used for the nodes. In the experiment, each node transmitted randomly selected DR9 packets with small random gaps between consecutive packets, where the maximum gap length was 100 ms. DR9 was used rather than DR8 to maximize the overall traffic load. All packets from the same node were transmitted with the same power. As the packet traces were collected with the same frequency group, a random group was selected for each packet by applying a frequency offset to the packet trace. The carrier frequency was 3.515 GHz and the sample rate was 500 kHz. The gateway simply received the signal from all nodes for 10 seconds and write the signal to a file to be processed. Fig.~\ref{fig:powderexpsig} shows the received signal by the gateway and the estimated signal from each individual node in a typical experiment, where the estimated signal from a node was the reconstructed signal of all decoded packets from this node calculated by the SIC component of the receiver. It can be seen that the nodes were transmitting packets almost non-stop and had very different signal powers. The signal power of the same node may also differ in different header and data fragments because of frequency-selective fading. Fig.~\ref{fig:powderexpres} shows the Packet Receiving Ratios (PRR) of all nodes, which is the average of 3 runs. It can be seen that the gateway was able to decode almost all packets from the nodes.

The experiment serves the important purpose of demonstrating our receiver design and the decoding of real-world signals that were received over long-distance links in an environment filled with interference. It was however beyond our capability to set up even larger networks to probe the limit of LR-FHSS. Also, currently, testbeds with real satellite links are not available. We therefore further evaluate LR-FHSS with trace-driven simulations, as discussed in the next section.

\section{Evaluation of LR-FHSS}
\label{sec:evaluation}

In this section, we report our findings of the performance of LR-FHSS obtained with trace-driven simulations.  

\subsection{Channel Models}

Three types of channels were tested, namely, the Additive White Gaussian Noise (AWGN) channel, the ETU channel~\cite{Ref3GPPTS36101,Ref3GPPTS36104}, and the NTN channel~\cite{3GPPTR38811,3GPPTR38901,3GPPTR3821,ITURP68111,Refmatlabntt}. The AWGN channel is the simplest where only white Gaussian noise was added to the signal. The ETU channel represents challenging terrestrial channels with strong multi-path, large channel fluctuation, and delay spread around 5~$\mu$s. The NTN channel represents non-terrestrial channels of satellite links. The number of antennas at the receiver was 1 for the AWGN channel and 2 for other channels, because the AWGN channel does not have antenna diversity. For the NTN channel, the carrier frequency was 10 GHz, the satellite speed was 7562.2 m/s, which were selected according to typical conditions of Low Earth Orbit (LEO) satellite links. The delay profile was {\fontfamily{qcr}\selectfont NTN-TDL-D}, which has a strong line-of-sight path, because most users will likely set up the satellite link with a clear view of the sky. It was also assumed that the node could estimate the Doppler frequency shift and approximate it as a constant, which can be achieved by analyzing downlink signals from the satellite, so that it could cancel most of the Doppler shift before transmitting any packet, 

\subsection{SNR Threshold of One-to-One Links}
\label{sec:singlepktres}

The SNR threshold is defined as the minimum SNR to maintain the PRR 0.9 or above, when only one packet is transmitted so that packet loss is not due to collision. It is an important metric that reveals the capability to deal with weak signals in one-to-one links. It may be worth mentioning that the SNR threshold has not been reported in earlier work except in the document by Semtech~\cite{semtechperformance}, because the measurement requires the decoding of actual LR-FHSS packets.

\begin{figure} 
	\begin{center}
		\includegraphics[width=\resfigw]{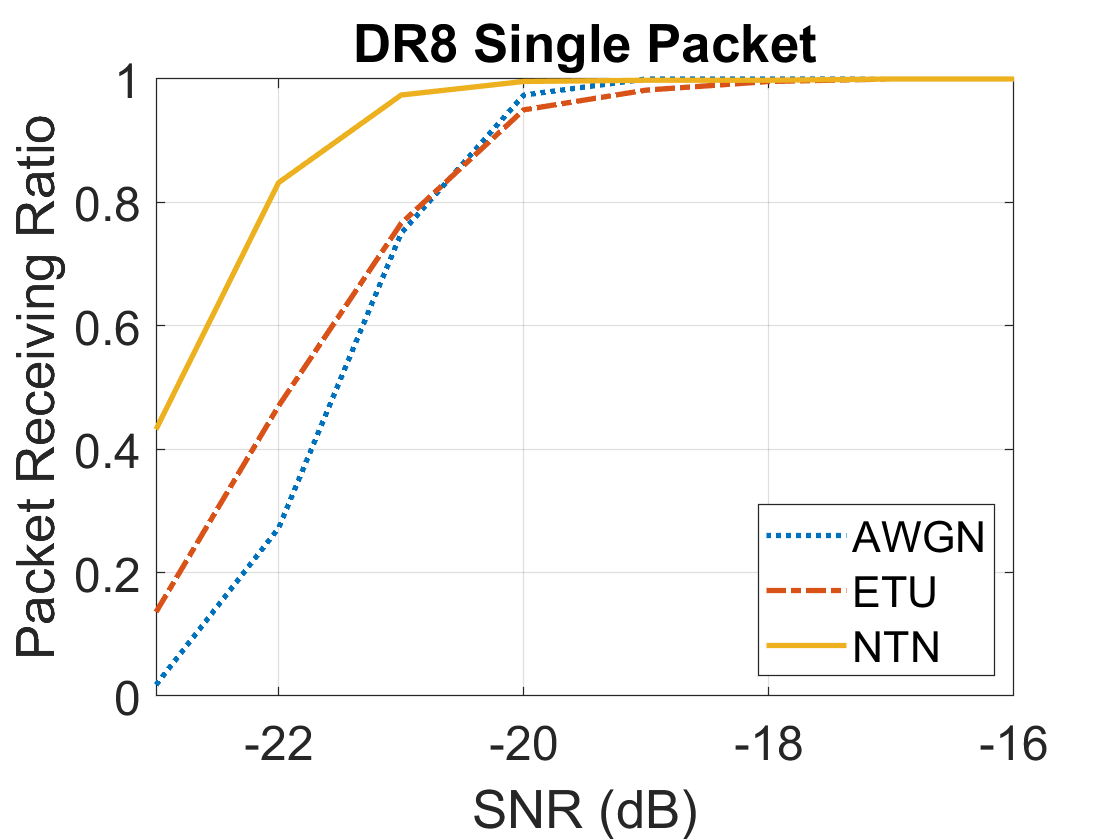} 
		\includegraphics[width=\resfigw]{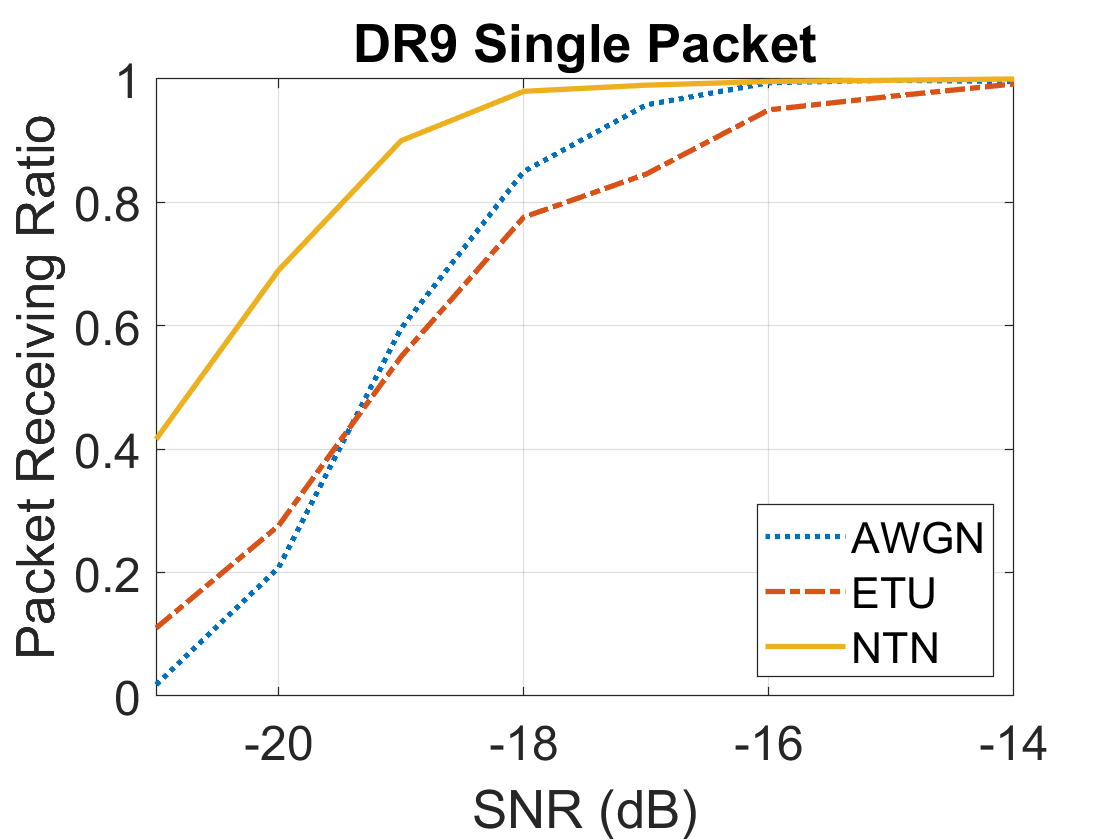} 
	\end{center}
	\vspace{\figspacesub}
	\caption{Receiving ratio of packets in one-to-one links.}
	\vspace{-0.1in}
	\label{fig:res_singlepkt}
\end{figure}

We synthesized signal traces by adding only one packet to the trace and scale it according to the SNR level. Fig.~\ref{fig:res_singlepkt} shows the results of DR8 and DR9 in all three types of channels. It can be seen that the SNR thresholds of DR8 are -20 dB, -20 dB, -21 dB for the AWGN, ETU, and NTN channels, respectively, while those of DR9 are -17 dB, -16 dB, and -19 dB, respectively. The SNR thresholds are similar to those of the Chirp Spread Spectrum (CSS) modulation under similar settings according to~\cite{LoRaCap}. Therefore, LF-FHSS should achieve a similar communication distance. ETU channel is the most challenging because of strong multi-path and large channel fluctuations. It is somewhat unexpected that tpe performance is the best in the NTN channel, which is because the {\fontfamily{qcr}\selectfont NTN-TDL-D} delay profile is dominated by a strong line-of-sight path and at the same time can benefit from antenna diversity.

\subsection{Network Capacity}

Network capacity is defined as the total number of data bits received by the gateway per second, under the constraint that the PRR is 0.9 or above. LR-FHSS was proposed mainly to improve the network capacity.

Our receiver is capable of processing composite signals from multiple nodes. We synthesized traces that contained multiple packets by adding signals of individual packets in the time-domain. 
To be more specific, a trace lasted 10 seconds and was initially only noise of unit power. Then, depending on the traffic load, a certain number of packets were randomly selected and added to the trace at random times. For each packet, first, a frequency group was randomly selected, according to which a frequency offset was applied to the packet. The packet was then passed through the channel, scaled according to its SNR, then added to the trace. The SNR was randomly selected within $[-15, 5]$ dB and $[-12, 2]$ dB for DR8 and DR9 packets, respectively. There was a range of 20 dB for both DR8 and DR9 to model errors in transmission power control. To focus on network-wide issues, the minimum SNR was a few dBs higher than the SNR threshold found in Section~\ref{sec:singlepktres}, so that packet loss was mainly caused by collision. 

\subsubsection{DR8 or CR9 Only}

\begin{figure} 
	\begin{center}
		\includegraphics[width=\resfigw]{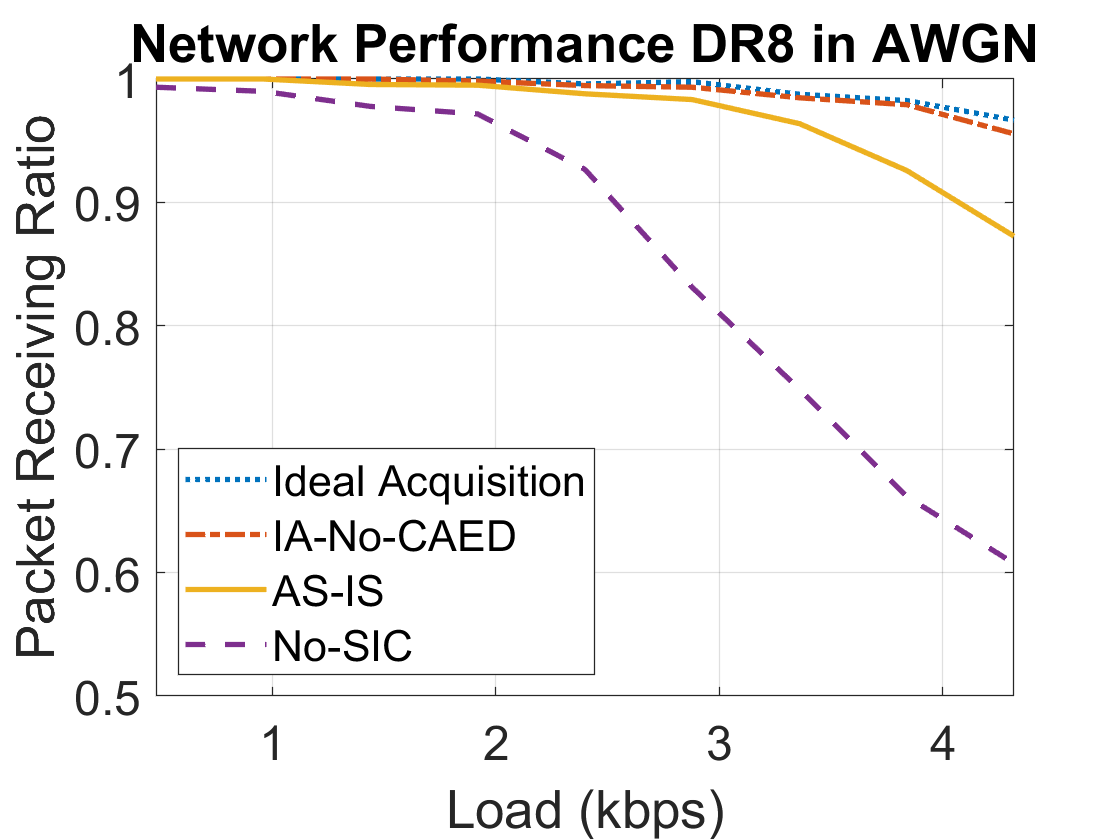} 
		\includegraphics[width=\resfigw]{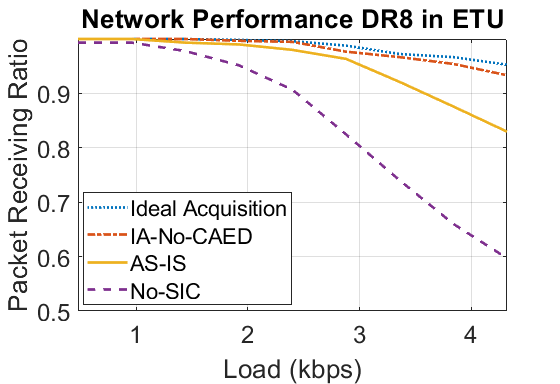} \\
		\includegraphics[width=\resfigw]{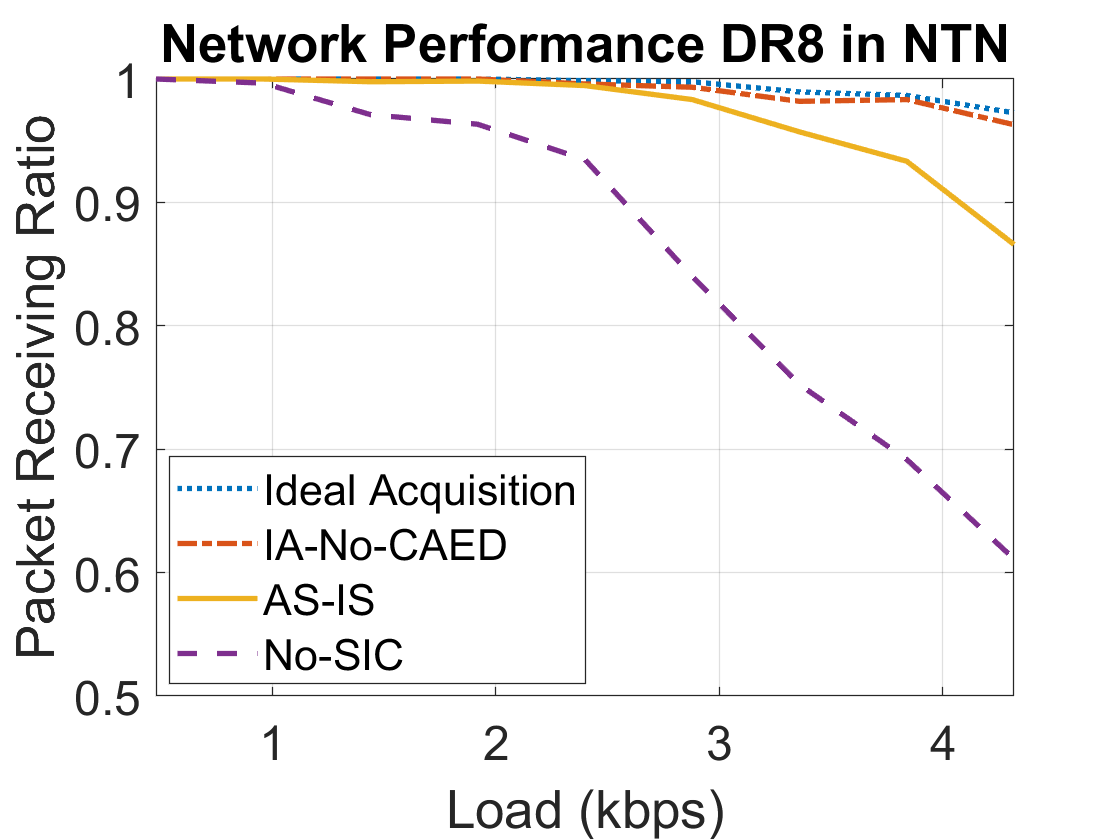} 
	\end{center}
	\vspace{\figspacesub}
	\caption{Network capacity with DR8.}
	\vspace{-0.1in}
	\label{fig:res_network_dr8}
\end{figure}

\begin{figure} 
	\begin{center}
		\includegraphics[width=\resfigw]{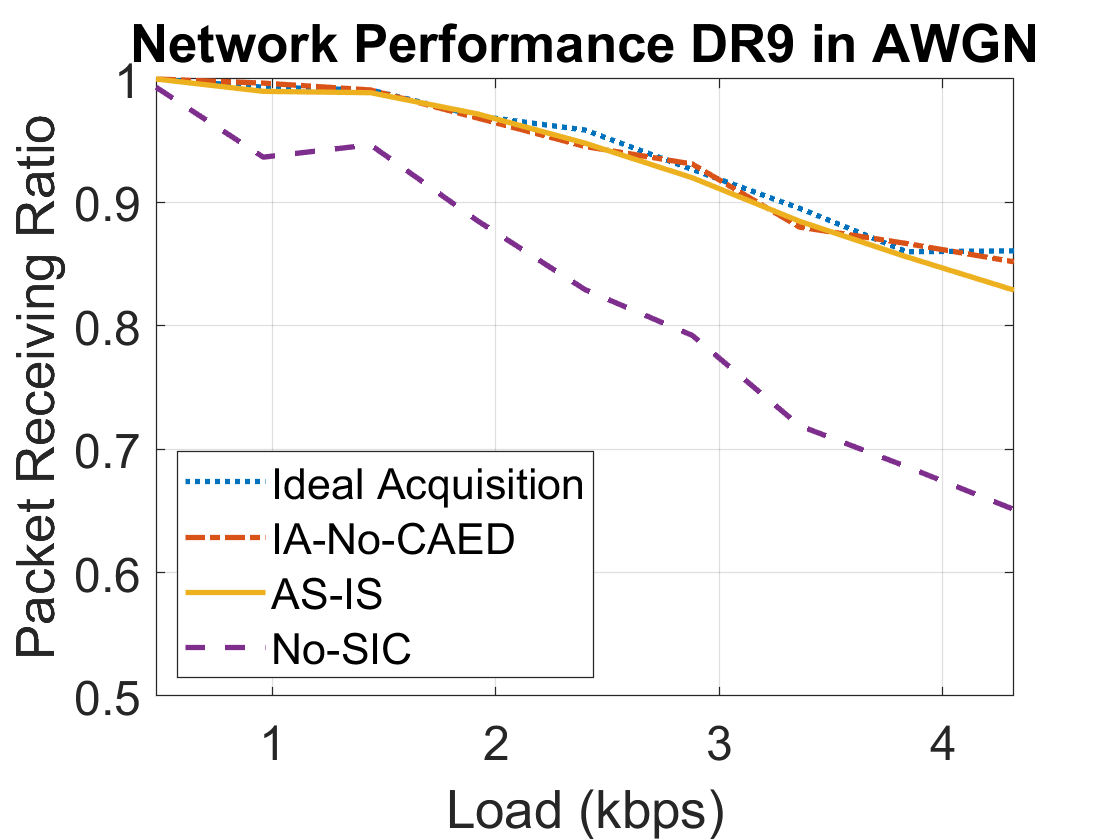} 
		\includegraphics[width=\resfigw]{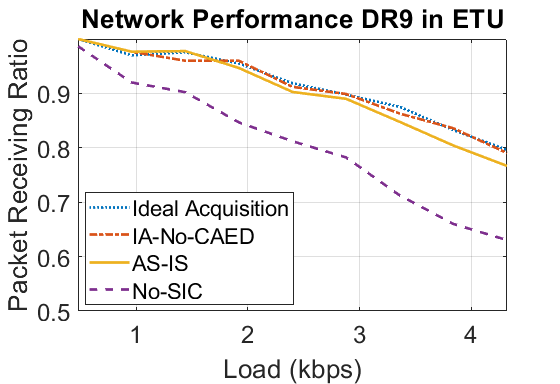} \\
		\includegraphics[width=\resfigw]{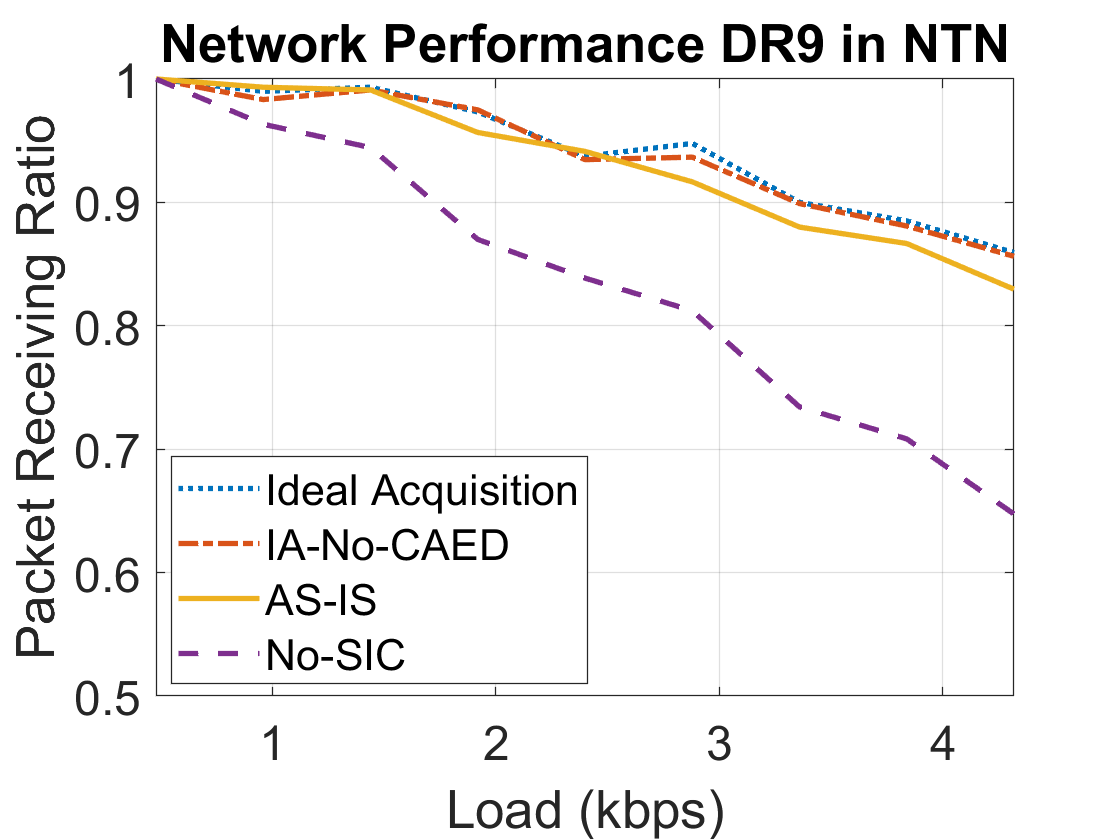} 
	\end{center}
	\vspace{\figspacesub}
	\caption{Network capacity with DR9.}
	\vspace{-0.1in}
	\label{fig:res_network_dr9}
\end{figure}

First, we synthesized traces consisting of packets with only one data rate. Fig.~\ref{fig:res_network_dr8} and  Fig.~\ref{fig:res_network_dr9} show the performance of DR8 and DR9, respectively. In the figures, ``AS-IS'' refers to the actual receiver we designed, while others refer to the actual receiver with certain variations for comparison. To be more exact, ``Ideal Acquisition'' refers to a receiver with perfect header acquisition, i.e., the  receiver knows the start time and frequencies of the packet and only needs to decode the data fragments. ``IA-No-CAED'' refers to ``Ideal Acquisition''  with CAED disabled. ``No-SIC'' refers to ``AS-IS'' without SIC. 

It can be seen that the network capacities of DR8 are 3.52 kbps, 3.22 kbps, and 3.51 kbps in AWGN, ETU, and NTN channels respectively, while those of DR9 are 2.74 kbps, 2.25 kbps, and 2.73 kbps, respectively. 
The network capacities for the same data rate are similar in all channels because packet loss is mainly caused by collisions if the SNR is sufficiently high. In~\cite{9422331}, it has been reported that the maximum network goodput is about 2 kbps. The results in~\cite{9653679} suggest that the network capacity is about 2 kbps. In~\cite{semtechperformance}, Semtech reported that the network capacity is 700,000 pkt/day, which is 8.1 pkt/sec. Although the exact size of the packets used in the evaluation was not given, the typical packet size discussed in~\cite{semtechperformance} was 20 bytes, suggesting a capacity of 1.3 kbps. The capacity achieved by our receiver is therefore consistently higher than those reported earlier in all channel models.

Our receiver achieves a higher network capacity because of more advanced signal processing techniques such as CAED and SIC. In fact, in all cases, there exists a large gap between ``No-SIC'' and ``AS-IS,'' confirming the importance of SIC. For DR8, there is a small, albeit noticeable, difference between IA-No-CAED and Ideal Acquisition, confirming the usefulness of CAED. For DR9, CAED is less helpful because the coding rate in DR9 is too high. The gaps between AS-IS and Ideal Acquisition are small when the traffic load is not high, suggesting that our header acquisition component works well even with collisions. The gaps are larger with DR8 because the packet transmission time with DR8 is much longer, leading to more collisions. DR8 however still achieves a higher capacity, which will be discussed further shortly.  


\subsubsection{Both DR8 and CR9}

\begin{figure} 
	\begin{center}
		\includegraphics[width=\resfigw]{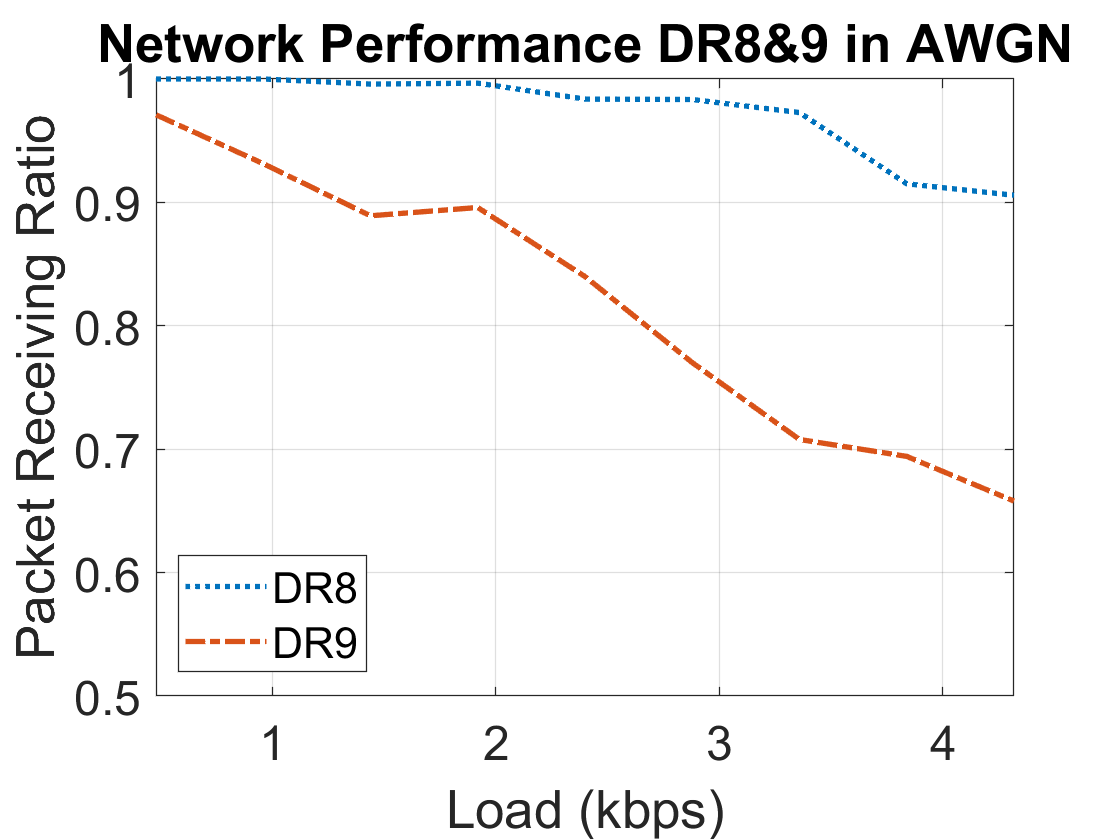} 
		\includegraphics[width=\resfigw]{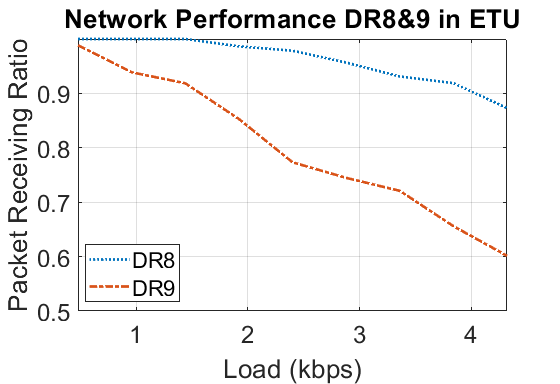} \\
		\includegraphics[width=\resfigw]{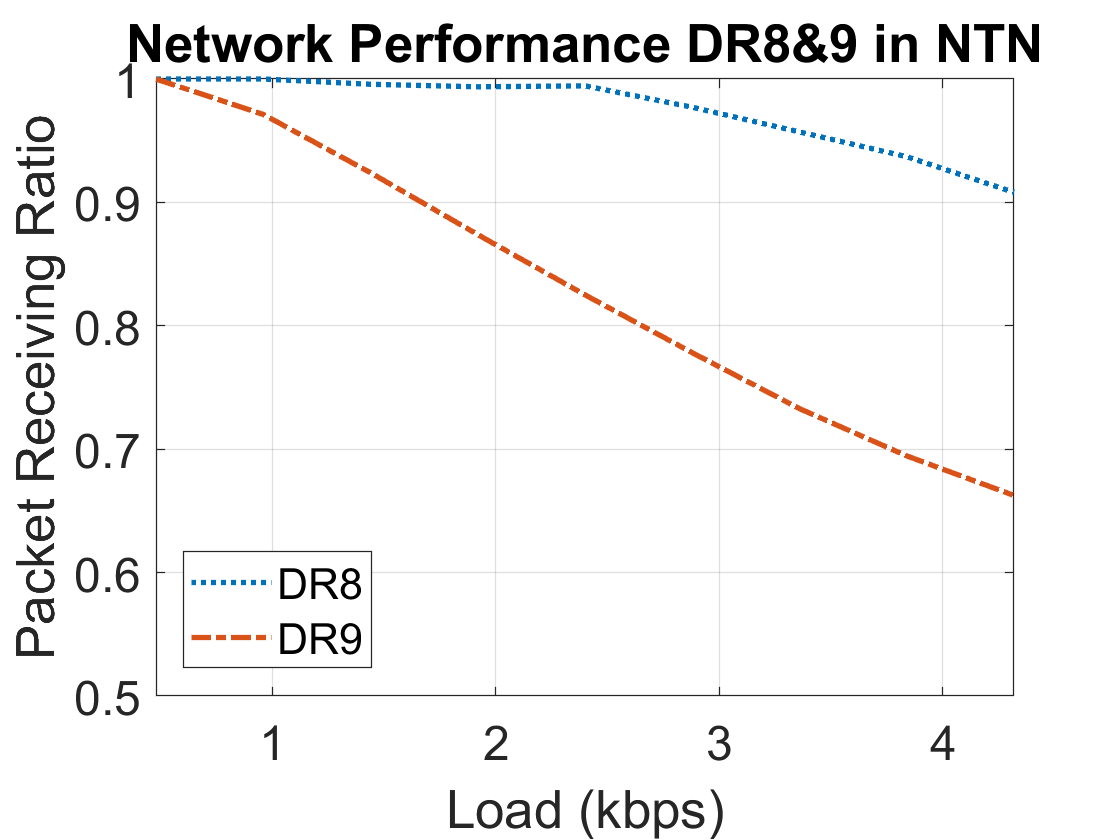} 
	\end{center}
	\vspace{\figspacesub}
	\caption{Network capacity with DR8 and DR9.}
	\vspace{-0.1in}
	\label{fig:res_network_drboth}
\end{figure}

We also synthesized traces when both DR8 and DR9 packets were randomly selected. Fig.~\ref{fig:res_network_drboth} shows the performance of AS-IS for DR8 and DR9, where it can be seen that there exists a large gap between DR8 and DR9, even when the SNR of DR8 was 3 dB lower than DR9 in the simulation. This could complicate rate selection because rate selection is usually determined by the channel condition. That is, a node should choose a higher data rate if its channel is stronger; and, with correct choices of data rates, all nodes should eventually enjoy roughly the same PRR. Therefore, some future research might be needed on rate selection for LR-FHSS.

\begin{figure} [h]
	\begin{center}
		\includegraphics[width=2in]{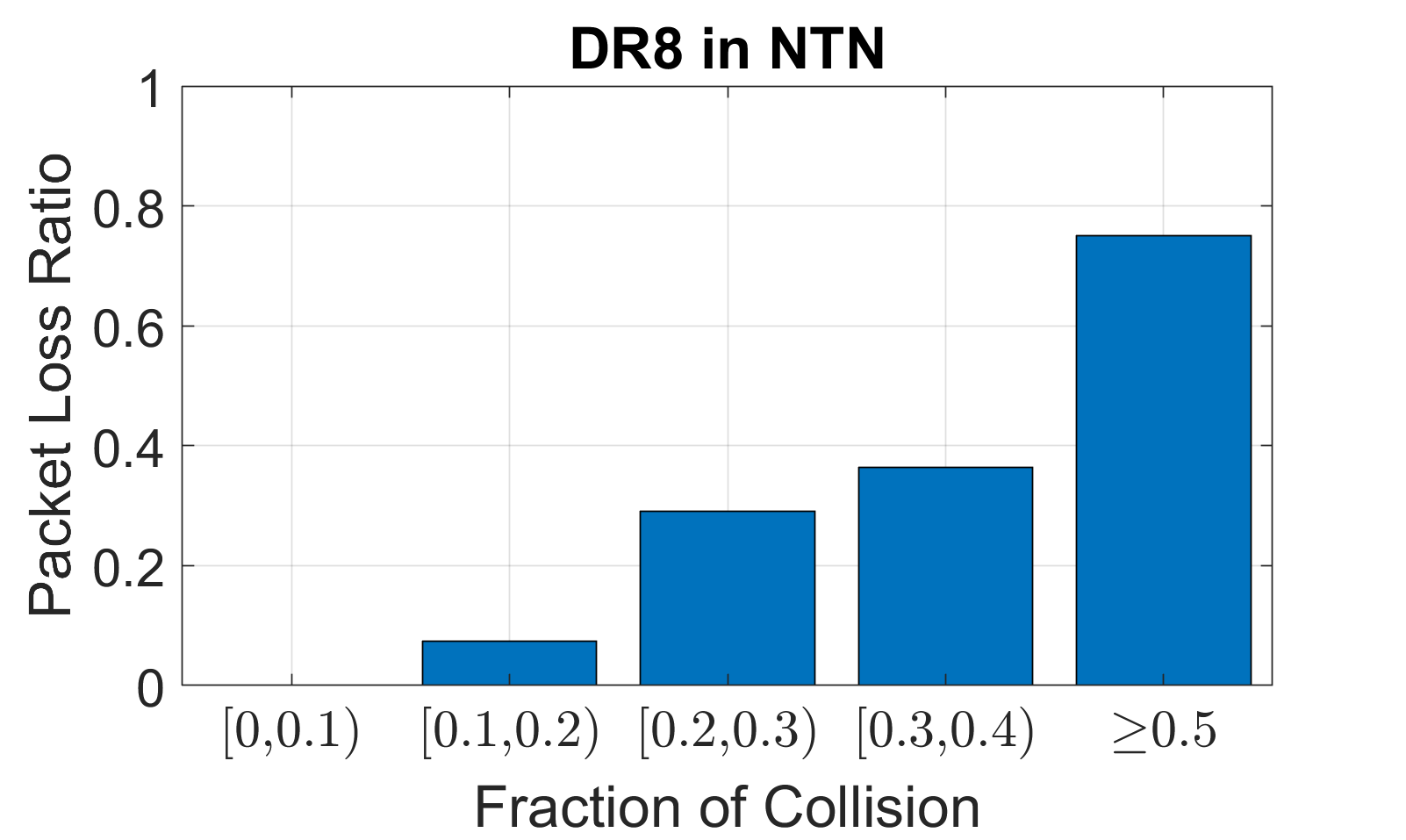} 
		\includegraphics[width=2in]{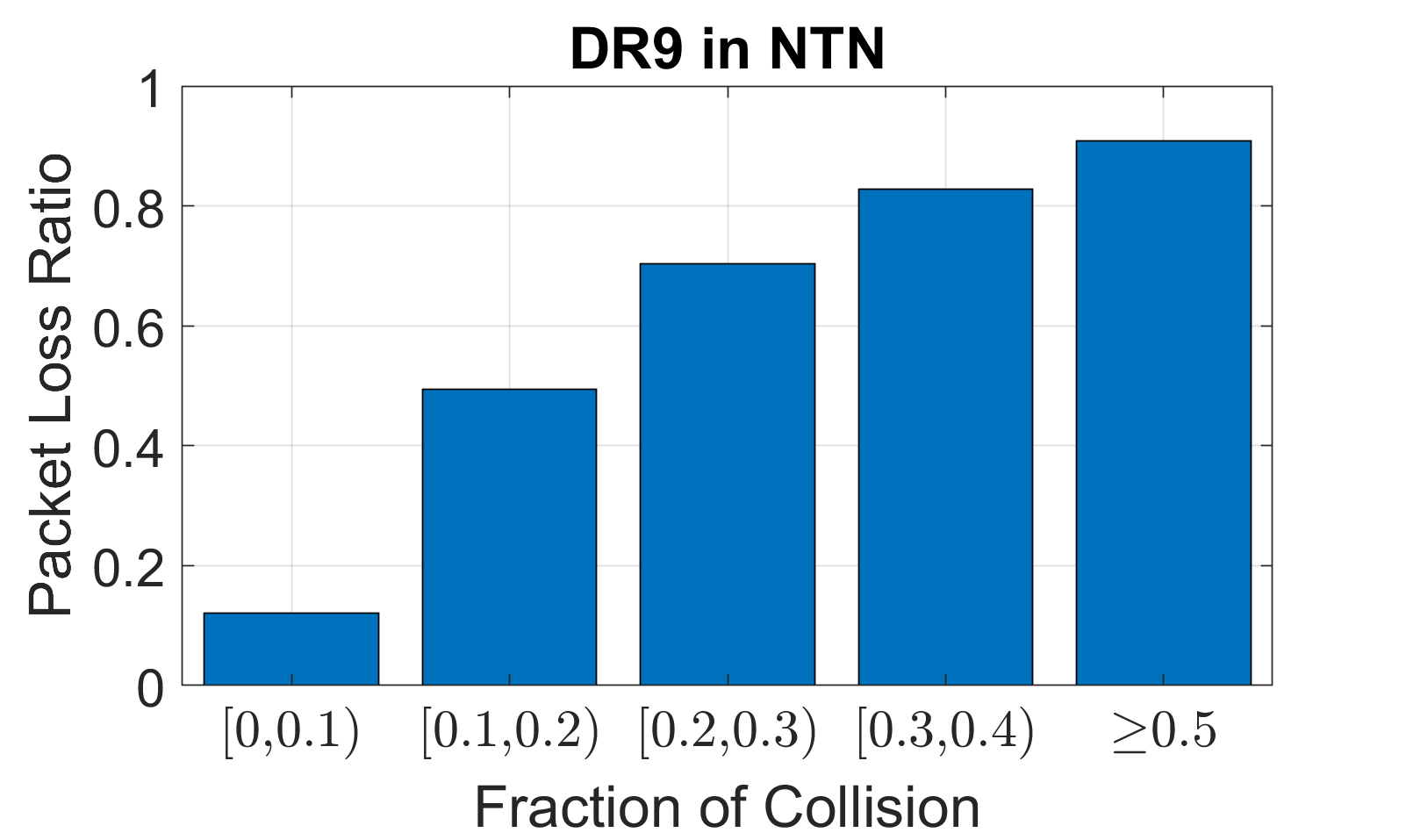} \\
		{\small (a)} \\
		\includegraphics[width=2in]{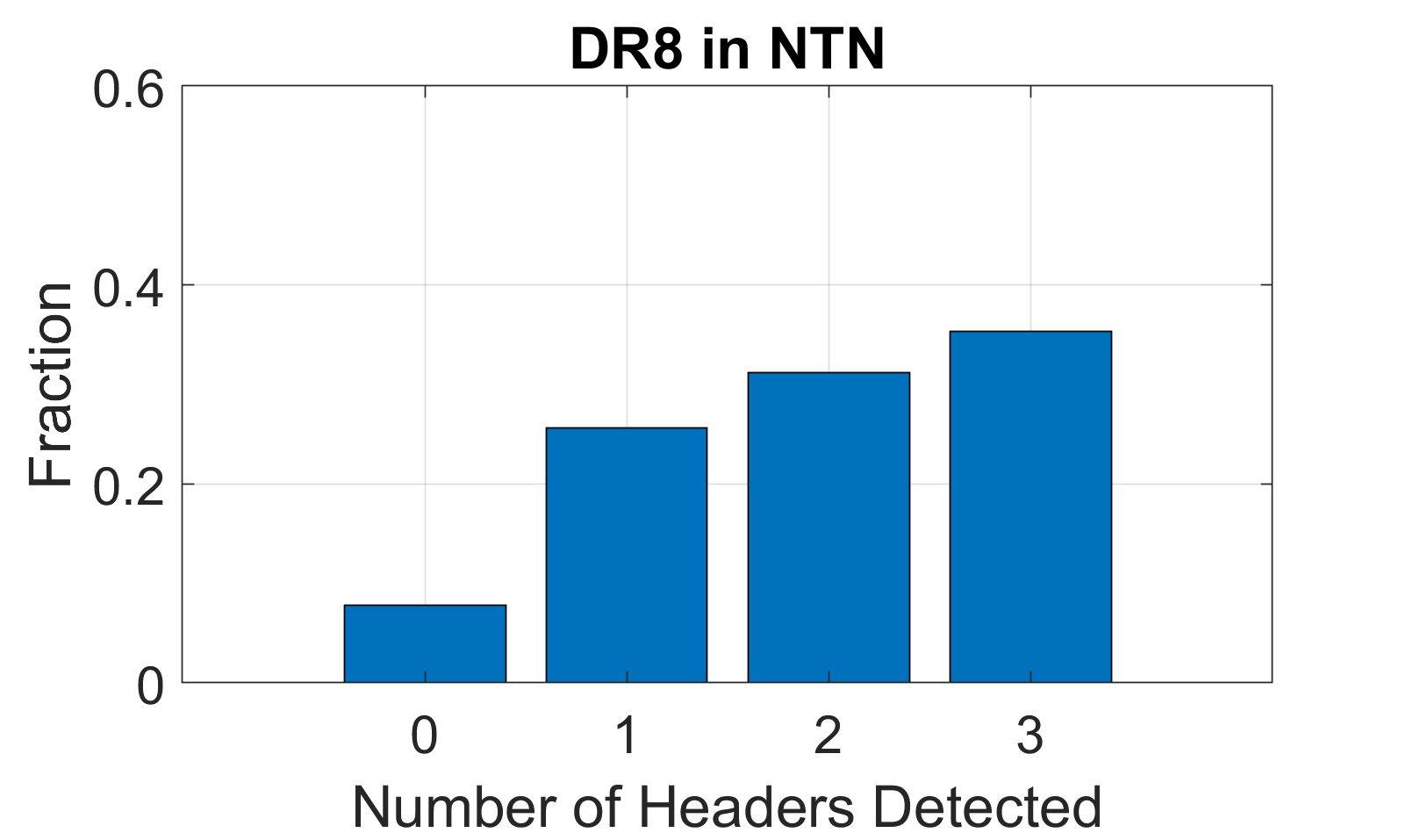} 
		\includegraphics[width=2in]{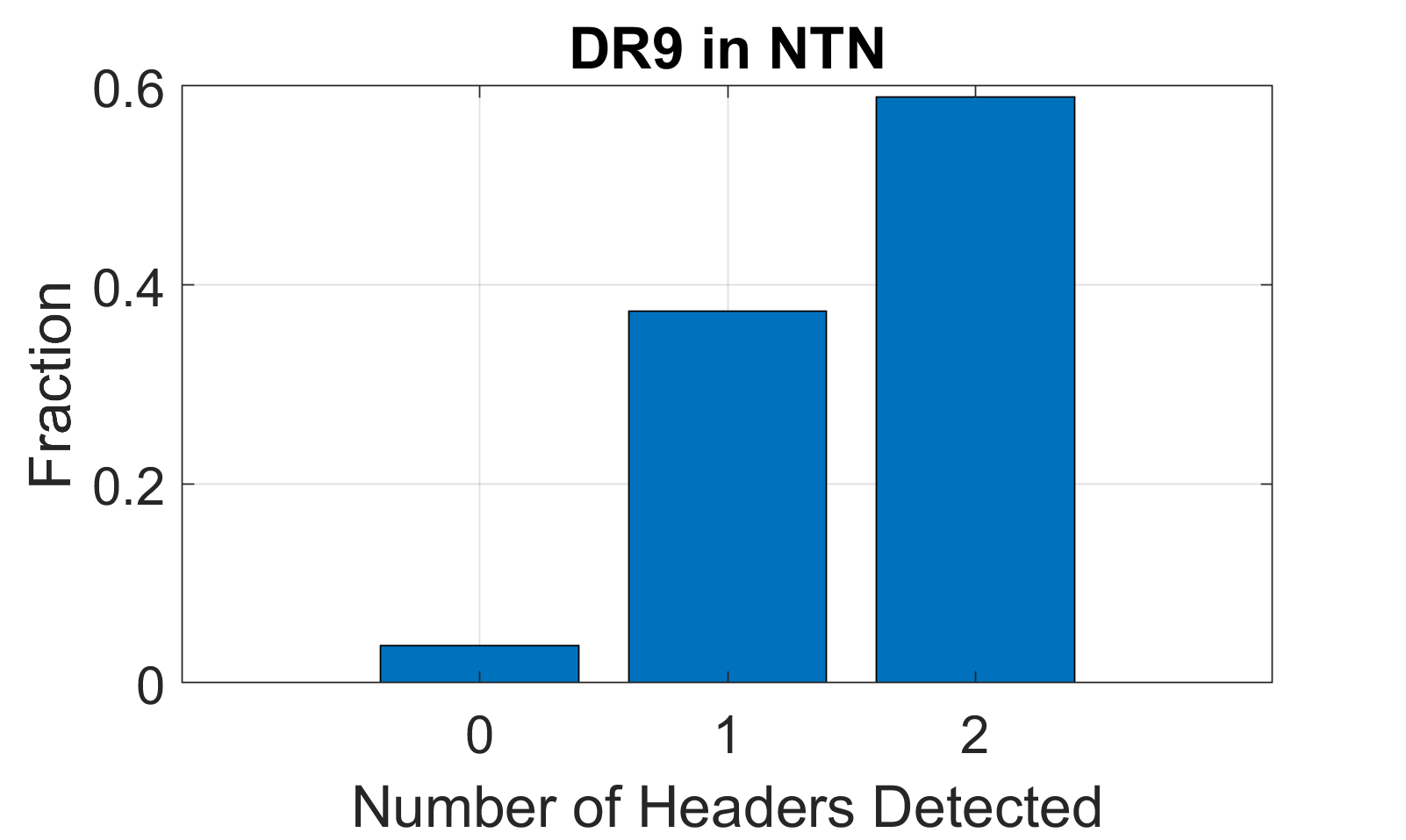} \\
		{\small (b)} \\
	\end{center}
	\vspace{\figspacesub}
	\caption{Further analysis when the load is 4.32 kbps. (a) Packet loss ratio as a function of collision level. (b) Number of headers detected.}
	\vspace{-0.1in}
	\label{fig:res_collevel}
\end{figure}

We found DR9 suffers more loss than DR8 mainly because, by design, it receives less protection from the error correction code. We examined the results under the highest traffic load. Due to the limit of space, we discuss the findings in the NTN channel because the general trends are the same for all channel models. Fig.~\ref{fig:res_collevel}(a) shows the loss ratio of packets as a function of the fraction of symbols under collision, where a symbol is considered under collision according to CAED. It can be seen that many packets with DR8 can still be received correctly even when a large fraction of symbols are under collision. For example, over 70\% of the packets can still be decoded correctly even when 20\% to 30\% of symbols are under collision. On the other hand, DR9 is much more sensitive to collision. To our surprise, under high traffic load, packet detection is not a main factor that leads to more packet loss with DR9. Fig.~\ref{fig:res_collevel}(b) shows the distributions of the number of headers detected for each packet. It can be seen that there were more packets with 0 header detected with DR8 than with DR9; in other words, actually, more packets with DR8 were lost due to failures of packet detection, even though DR9 transmits one less header replica. We believe this is because headers in DR8 and DR9 use exactly the same format and code rate, while fewer transmitted header replicas with DR9 is compensated for by its higher signal power. 

\section{Conclusions}
\label{sec:conclusion}

LR-FHSS is an important addition to the LoRa family. Preferably, it should be evaluated in a setting as close to the real-world as possible, which was difficult because there was no sufficient open documentations. We started this work facing many challenges, such as obtaining the complete understanding about the physical layer modulation and designing a receiver from scratch to convert the baseband waveform into bits. We were able to overcome these challenges and perform trace-driven simulations to reveal the performance of LR-FHSS, where real-world packet signals were processed and practical issues such as timing error, frequency error, error correction, and interference cancellation were taken into account. We also designed customized methods for LR-FHSS to improve its performance. We have uploaded our source code and trace files with free access.

\newpage
\bibliographystyle{plain}
\bibliography{zvl,myref_MZC,lrfhss_ref,star_ref}

\end{document}
\endinput